\documentclass{ieeeaccess}
\usepackage{cite}
\usepackage{amsmath,amssymb,amsfonts}
\usepackage{algorithmic}
\usepackage{graphicx}
\usepackage{textcomp}
\usepackage{braket}
\usepackage{caption}

\usepackage{url}
\usepackage[colorlinks]{hyperref}
\hypersetup{%
	plainpages=true,
	breaklinks=true,       
	hypertexnames=false,  
	pageanchor=true,
	colorlinks=true,
	linkcolor={black},
	citecolor={black},
	urlcolor={black},
	anchorcolor={black}
}

\def\BibTeX{{\rm B\kern-.05em{\sc i\kern-.025em b}\kern-.08em
    T\kern-.1667em\lower.7ex\hbox{E}\kern-.125emX}}
\begin{document}
\doi{10.1109/TQE.2020.DOI}

\title{Sensing of Low-Frequency Electric Fields Using Rydberg EIT within the Fisher Information Framework}
%
%

\author{
\uppercase{Tianyu Zhou}\authorrefmark{1},
\uppercase{Haipeng Xie}\authorrefmark{1},
\uppercase{Xin Wang}.\authorrefmark{2}
}

\address[1]{School of Electrical Engineering, Xi'an Jiaotong University, Xi'an 710049, China}
\address[2]{Shaanxi Province Key Laboratory of Quantum Information and 
	Quantum Optoelectronic Devices, School of Physics, Xi'an Jiaotong 
	University, 
	Xi'an 710049, People's Republic of China}

\corresp{Corresponding authors: Haipeng Xie (email: haipengxie@xjtu.edu.cn ) and Xin Wang (email: wangxin.phy@xjtu.edu.cn)}

\begin{abstract}

Rydberg atoms, which possess exceptionally large electric dipole moments, offer a promising route for electric field sensing as well as metrology traceable to the International System of Units (SI); however, current research predominantly focuses on the microwave (MW) regime, leaving the quasi-direct current (quasi-DC) and low-frequency bands, ubiquitous in power systems, largely unexplored. In this paper, we present a theoretical investigation into low-frequency electric field detection. To this end, we establish a comprehensive modeling framework incorporating Fisher information (FI) and the Cram\'{e}r-Rao lower bound (CRLB) to quantify the fundamental precision limits of electromagnetically induced transparency (EIT) readouts. Building upon this framework, we propose a linearized sensing strategy utilizing a DC-biased two-point differential measurement. Numerical validations demonstrate that this approach effectively mitigates the weak-field insensitivity for both DC and AC fields, achieving a CRLB-limited sensitivity bound of approximately $1\times 10^{-4}$ V/m/$\sqrt{\text{Hz}}$. Furthermore, to surpass the single-pass sensitivity limit, we introduce a Fabry-P\'{e}rot (FP) cavity-enhanced configuration. This architecture leverages intracavity phase modulation to significantly steepen the transmission slope, boosting the FI by over two orders of magnitude compared to standard free-space configurations. This work provides a rigorous theoretical basis and design guidance for the high-precision quantum monitoring of electromagnetic environments in smart grids.

\end{abstract}


\titlepgskip=-15pt

\maketitle

\section{Introduction}
\label{sec:introduction}
\PARstart{A}{s} modern electric power systems evolve toward higher voltage levels, larger scale interconnection, and pervasive digital intelligence, the electromagnetic environment surrounding power apparatus becomes increasingly complex\cite{Farhangi2010,Amin2005,Liu2014}. The electric field, as a direct physical observable that reflects both ambient electromagnetic conditions and the operating state of equipment, plays a foundational role in insulation condition assessment, fault early warning, and electromagnetic-environment monitoring\cite{Kuffel2000}. Conventional engineering approaches to electric field measurement are mainly based on (i) electrical sensors that infer field strength from induced charge or displacement current\cite{Baum1978,Riehl2003}, and (ii) optical sensors that convert an external field into phase (or intensity) modulation via linear/second-order electro-optic effects (Pockels/Kerr)\cite{Hidaka1989,Chmielak2011,Shimizu1996,Qiu2000}. In practice, however, intrinsic device noise and environmental perturbations commonly lead to offset drift, limited long-term stability, and constrained immunity to electromagnetic interference\cite{Li2025,Zeng2012,Li2013}. These limitations make it difficult for traditional solutions to simultaneously meet the more stringent requirements of future smart grids in terms of resolution, dynamic range, and reliability, especially under harsh high-voltage conditions\cite{Tzelepis2020}.

Against this backdrop, advances in laser frequency stabilization, coherent control of atoms, and quantum measurement have established quantum sensing as a promising and potentially SI-traceable approach to electromagnetic-field measurement\cite{Degen2017,Pirandola2018,Giovannetti2011,Cronin2009,Budker2007,Taylor2008,Kitching2018}. In particular, Rydberg atoms possess exceptionally large electric dipole moments and polarizabilities, resulting in extreme sensitivity to external electric fields and thereby offering an attractive platform for high-precision field sensing\cite{Mohapatra2008,Sedlacek2012,Kumar2017,Jing2020,Zhang2023,Wu2023,Ding2022,Liu2022,Wang2000,Liang2025,Peng2018,Liu2025}. In the Rydberg-atom electromagnetically induced transparency (EIT)\cite{Boller1991,marangos1998,Fleischhauer2005} framework, microwave (MW) electric field measurement was first demonstrated through absolute calibration based on spectral readout, and subsequent progress, including modulation spectroscopy, has achieved sensitivities on the order of $\mu\mathrm{V/cm}/\sqrt{\text{Hz}}$\cite{Sedlacek2012}. More recently, receiver-like architectures (e.g., superheterodyne concepts) have further pushed the sensitivity toward the $\mathrm{nV/cm}/\sqrt{\text{Hz}}$ regime\cite{Jing2020}, while also clarifying the scaling with atom number and the impact of non-idealities. In parallel, strategies such as critical enhancement\cite{Ding2022}, decoding via deep learning\cite{Liu2022}, and cavity enhancement\cite{Wang2000,Liang2025,Peng2018,Liu2025} have been explored to improve weak-signal detectability and robustness. Nevertheless, existing studies predominantly target the MW band and have demonstrated early potential in a few application domains (e.g., wireless communications, imaging, and device testing)\cite{Wu2025,Mao2024}. However, for power-system scenarios, quantum measurement of low-frequency (DC--kHz) electric fields, in particular power-frequency fields\cite{Han2025,Song2024,Arumugam2025,Ma2020,Lei24,Diao2025,Li23,Jau2020,Duspayev2024,Yan2025,Xiao2025,Xiao2024}, remains at an early exploratory stage. A central challenge is that low-frequency fields interact with Rydberg atoms through the DC Stark effect, which shifts the EIT resonance by an amount proportional to the square of the field amplitude. This quadratic dependence causes the response slope to approach zero in the weak-field limit, thereby limiting the readout sensitivity. Furthermore, conventional peak-shift readout requires the Stark-induced spectral displacement to exceed the full width at half maximum (FWHM) of the EIT resonance, which imposes a relatively high minimum detectable field threshold. These physical constraints indicate that key elements---including suitable theoretical models, linearized readout mechanisms, and engineering implementation pathways---are not yet mature and call for systematic investigation.

To address this gap, this paper focuses on Rydberg-atom quantum measurement of power-frequency electric fields and establishes a corresponding modeling and signal-readout framework. Specifically, we introduce Fisher information (FI) and the Cram\'{e}r-Rao lower bound (CRLB) into the analysis of EIT spectra, creating a quantitative mapping from spectral response to estimation uncertainty and ultimately to electric field sensitivity, enabling predictive assessment of achievable performance. To overcome the common loss of readout sensitivity in the weak-field regime caused by quadratic response, we propose and employ a linearized strategy based on a bias electric field combined with two-point differential readout, thereby enhancing identifiability in the weak-field limit. Finally, we introduce a Fabry-P\'{e}rot (FP) cavity-enhanced configuration that theoretically enhances the FI by over two orders of magnitude and correspondingly improves the detection sensitivity by more than one order of magnitude, thereby providing concrete design guidance for the subsequent engineering realization of highly precise sensors for power-frequency electric fields.

\begin{figure}
	\centering  
	\includegraphics[width=0.85\linewidth]{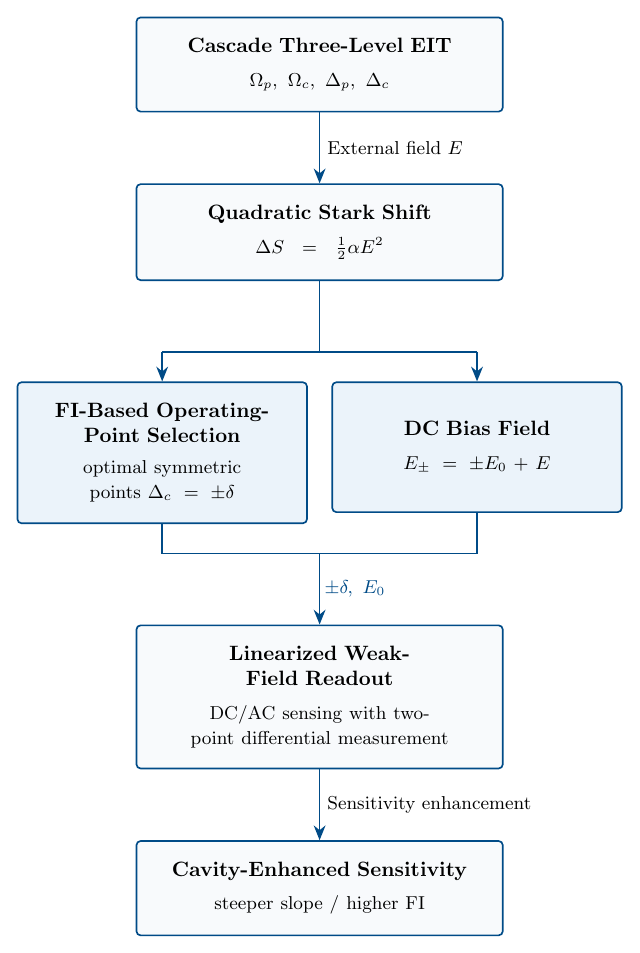}
	\caption{Schematic overview of the proposed low-frequency electric field sensing framework. In the cascade three-level EIT system, an external electric field induces a quadratic Stark shift of the resonance. Based on this response, FI-based operating point selection identifies the optimal symmetric detunings, while a DC bias field is introduced to enable linearization in the weak-field regime. These two elements together lead to a linearized readout for DC and AC electric field sensing, which can be further improved by cavity enhancement.}
	\label{fig1}
\end{figure}

The overall framework of the proposed sensing scheme is summarized in Fig.~\ref{fig1}. In the cascade three-level EIT system, an external electric field induces a quadratic Stark shift of the resonance. Based on this response, FI-based operating point selection determines the optimal symmetric detunings, while a DC bias field enables weak-field linearization. Together, these two elements realize a linearized readout for DC and AC electric field sensing. Cavity enhancement is further introduced to improve the achievable sensitivity.

\section{Three-Level Rydberg-EIT Model and Sensing Principle}

This section outlines the physical principles of Rydberg-based electric field sensing. Section~\ref{subsec:EIT_model} introduces the three-level cascade model and the probe transmission spectrum, while Section~\ref{subsec:AT_Stark} uses MW-induced Autler-Townes splitting (ATS)~\cite{Autler1955,Abi2010} mainly for context and emphasizes the Stark shift~\cite{Stark1913} under low-frequency electric fields, which is the focus of this work.

\subsection{Three-Level Cascade EIT Model}
\label{subsec:EIT_model}

\begin{figure}
	\centering  
	\includegraphics[width=0.5\linewidth]{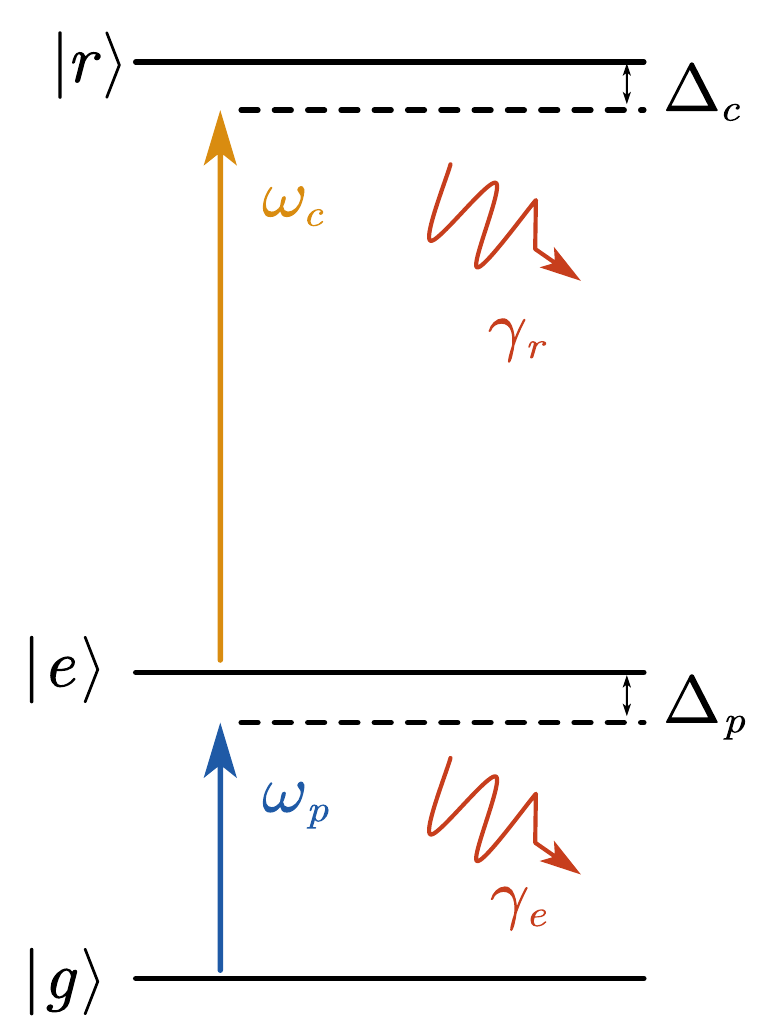}
	\caption{Cascade three-level scheme for EIT. The probe (blue) field at frequency $\omega_p$ couples the $\ket{g}\!\rightarrow\!\ket{e}$ transition with detuning $\Delta_p$, while the coupling (orange) field at frequency $\omega_c$ couples the $\ket{e}\!\rightarrow\!\ket{r}$ transition with detuning $\Delta_c$. The combined relaxation and decoherence effects for the intermediate and Rydberg states are represented by the effective rates $\gamma_e$ and $\gamma_r$, respectively (red wavy arrows).}
	\label{fig:cascade}
\end{figure}

We begin by introducing the basic concept of Rydberg EIT using a prototypical cascade three-level system. As illustrated in Fig.~\ref{fig:cascade}, the atom is modeled with three relevant states: the ground state $\ket{g}$, an intermediate excited state $\ket{e}$, and a high-lying Rydberg state $\ket{r}$. A weak probe field at frequency $\omega_p$ addresses the $\ket{g}\!\rightarrow\!\ket{e}$ transition (e.g., $5\mathrm{S}_{1/2}\!\rightarrow\!5\mathrm{P}_{3/2}$ in Rb or $6\mathrm{S}_{1/2}\!\rightarrow\!6\mathrm{P}_{3/2}$ in Cs), while a strong coupling field at frequency $\omega_c$ drives the $\ket{e}\!\rightarrow\!\ket{r}$ transition to a Rydberg level with principal quantum number typically with $n\approx 20$--$50$
, depending on the implementation. The associated single-photon detunings are
\begin{equation}
	\Delta_p = \omega_{eg}-\omega_p, \qquad
	\Delta_c = \omega_{re}-\omega_c,
\end{equation}
where $\omega_{eg}$ and $\omega_{re}$ are the transition frequencies of $\ket{g}\!\rightarrow\!\ket{e}$ and $\ket{e}\!\rightarrow\!\ket{r}$, respectively. In realistic systems, the atomic states are subject to population decay and environmental dephasing driven by the finite lifetime of the intermediate state, transit-time broadening, and laser phase noise. We model these combined relaxation and decoherence effects through the effective rates $\gamma_e$ and $\gamma_r$, which largely determine the EIT linewidth and contrast.

Under the rotating-wave approximation (RWA) and in a suitable interaction picture, the effective Hamiltonian of the three-level system takes the form
\begin{equation}
	H = \frac{\hbar}{2}
	\begin{pmatrix}
		0             & \Omega_p^{*} & 0 \\
		\Omega_p      & -2\Delta_p   & \Omega_c^{*} \\
		0             & \Omega_c     & -2(\Delta_p + \Delta_c)
	\end{pmatrix},
\end{equation}
where $\Omega_p = \mu_{ge} E_p/\hbar$ is the probe field Rabi frequency, $\mu_{ge}$ is the electric dipole matrix element for the $\ket{g}\!\rightarrow\!\ket{e}$ transition, and $E_p$ is the probe field amplitude; $\hbar$ denotes the reduced Planck constant. The coupling field Rabi frequency $\Omega_c$ is defined analogously.

The system dynamics are described by the density matrix
\begin{equation}
	\rho =
	\begin{pmatrix}
		\rho_{gg} & \rho_{ge} & \rho_{gr} \\
		\rho_{eg} & \rho_{ee} & \rho_{er} \\
		\rho_{rg} & \rho_{re} & \rho_{rr}
	\end{pmatrix},
\end{equation}
where the diagonal elements $\rho_{ii}$ $(i\in\{g,e,r\})$ denote the state populations and the off-diagonal elements $\rho_{ij}$ $(i\neq j)$ describe quantum coherences between $\ket{i}$ and $\ket{j}$. The time evolution of $\rho$ is governed by the Lindblad master equation 
\begin{equation}
	\dot{\rho}
	= -\frac{i}{\hbar}[H,\rho] + \mathcal{L}(\rho),
\end{equation}
where $[H,\rho]=H\rho-\rho H$, and $\mathcal{L}(\rho)$ is the Lindblad super-operator accounting for relaxation and decoherence. By solving the above equation in the steady state, one obtains the stationary density matrix of the system.

Since the probe transmission is of primary interest, we focus on the coherence element $\rho_{ge}$ associated with the $\ket{g}\!\rightarrow\!\ket{e}$ transition. For a cascade three-level system, the steady-state solution for $\rho_{ge}$ can be written as
\begin{equation}
	\rho _{ge}=\frac{i\Omega _p\bigl( \gamma _r+i2(\Delta _p+\Delta _c) \bigr)}{\bigl( \gamma _e+i2\Delta _p \bigr) \bigl( \gamma _r+i2(\Delta _p+\Delta _c) \bigr) +|\Omega _c|^2}.
\end{equation}
This coherence is directly related to the optical response of the medium: in the EIT regime, the imaginary part $\mathrm{Im}[\rho_{ge}]$ is proportional to the probe absorption, whereas the real part $\mathrm{Re}[\rho_{ge}]$ corresponds to the dispersive refractive index response. As shown by the numerical evaluation in Fig.~\ref{fig:EIT}, when the probe is held on resonance ($\Delta_p=0$), the solid black curve corresponding to the three-level EIT spectrum exhibits a pronounced and narrow transparency feature in the absorption signal $\mathrm{Im}[\rho_{ge}]$ as a function of the coupling detuning $\Delta_c$. This central peak occurs at the two-photon resonance condition $\Delta_p+\Delta_c=0$ (here $\Delta_c=0$) and reflects the formation of the EIT dark state via destructive quantum interference.

The optical response of the medium experienced by the probe field is characterized by the susceptibility, which is proportional to $\rho_{ge}$. In the weak probe limit, the complex susceptibility takes the form
\begin{equation}
	\chi
	= -\frac{2 N \mu_{ge}^2}{
		\varepsilon_0 \hbar \Omega_p
	}\,
	\rho_{ge},
	\label{chi}
\end{equation}
where $N$ is the atomic number density and $\varepsilon_0$ is the vacuum permittivity. According to the Beer--Lambert law, the transmitted probe field intensity after propagating through a vapor cell of length $l$ is given by
\begin{equation}
	\eta  = \eta_0 \exp\bigl[-k_p l\,\mathrm{Im}(\chi)\bigr],
	\label{eta}
\end{equation}
where $\eta_0$ is the incident probe field intensity and $k_p$ is the probe wave vector. This expression gives rise to the characteristic EIT transmission spectrum, featuring a narrow transparency peak around the two-photon resonance condition $\Delta_p+\Delta_c\simeq 0$.

The EIT spectrum thus serves as the reference signal for field sensing. External fields that couple to the Rydberg state modify the position and shape of the EIT window. By tracking these changes in the EIT spectrum, one can infer the strength of the applied field, which forms the basic operating principle of Rydberg atom field sensors.

\subsection{Basic Principles of Rydberg Atom Electric Field Sensing}
\label{subsec:AT_Stark}

\begin{figure}
	\centering  
	\includegraphics[width=\linewidth]{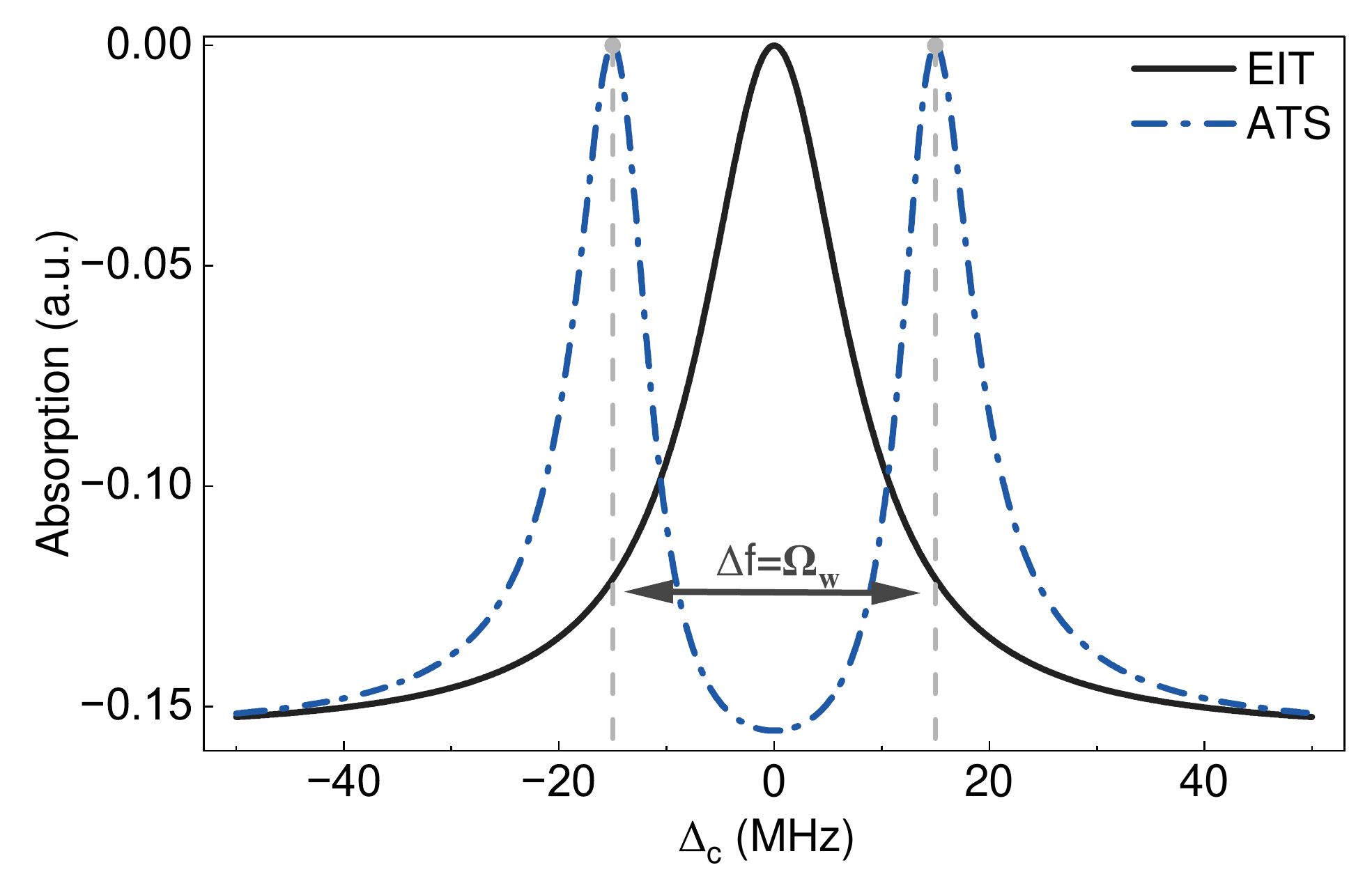}
	\caption{Simulated spectra (EIT and ATS) versus coupling detuning $\Delta_c$ at $\Delta_p=0$. The solid black curve represents the three-level EIT profile with $\Omega_p/2\pi=1~\mathrm{MHz}$, $\Omega_c/2\pi=10~\mathrm{MHz}$, and decay rates $\gamma_e/2\pi=6.066~\mathrm{MHz}$, $\gamma_r/2\pi=0.004~\mathrm{MHz}$. The dash-dotted blue curve depicts the four-level ATS under the same parameters but with an additional MW coupling field. The applied MW field induces a transparency window splitting denoted by $\Delta f = \Omega_w/2\pi = 30~\mathrm{MHz}$.}
	\label{fig:EIT}
\end{figure}

For MW fields, the well-established ATS readout applies: a resonant MW field couples two Rydberg states and splits the EIT peak into a spectral doublet (see the dash-dotted curve in Fig.~\ref{fig:EIT}), with the splitting directly yielding the field amplitude,
\begin{equation}
	E_s
	= \frac{\hbar \Delta f}{\sqrt{2}\,\mu_s},
\end{equation}
where $\Delta f$ is the measured peak separation and $\mu_s$ is the relevant electric dipole matrix element. This linear mapping underpins most existing Rydberg MW/RF sensors.

In this work, however, the target fields arise in power-system environments and typically lie at power-line frequencies or in the kHz range. These low-frequency fields are far below typical Rydberg--Rydberg transition frequencies and therefore cannot resonantly couple adjacent Rydberg levels, making ATS-based readout inapplicable in this context. Instead, low-frequency and DC electric fields are sensed through the Stark shift of a selected Rydberg state.

For DC fields, the Stark shift is well approximated by a quadratic dependence,
\begin{equation}
	\Delta S = \frac{1}{2}\,\alpha E^2,
	\label{Eq:Stark}
\end{equation}
where $\alpha$ is the polarizability of the Rydberg state and $E$ is the external electric-field amplitude. The polarizability can be obtained by diagonalizing the atomic Hamiltonian in the presence of a static electric field. In this work, the numerical evaluation is carried out using the open-source ARC package (Alkali.ne Rydberg Calculator)~\cite{Sibalic2017}.

\begin{figure}
	\centering
	\includegraphics[width=\linewidth]{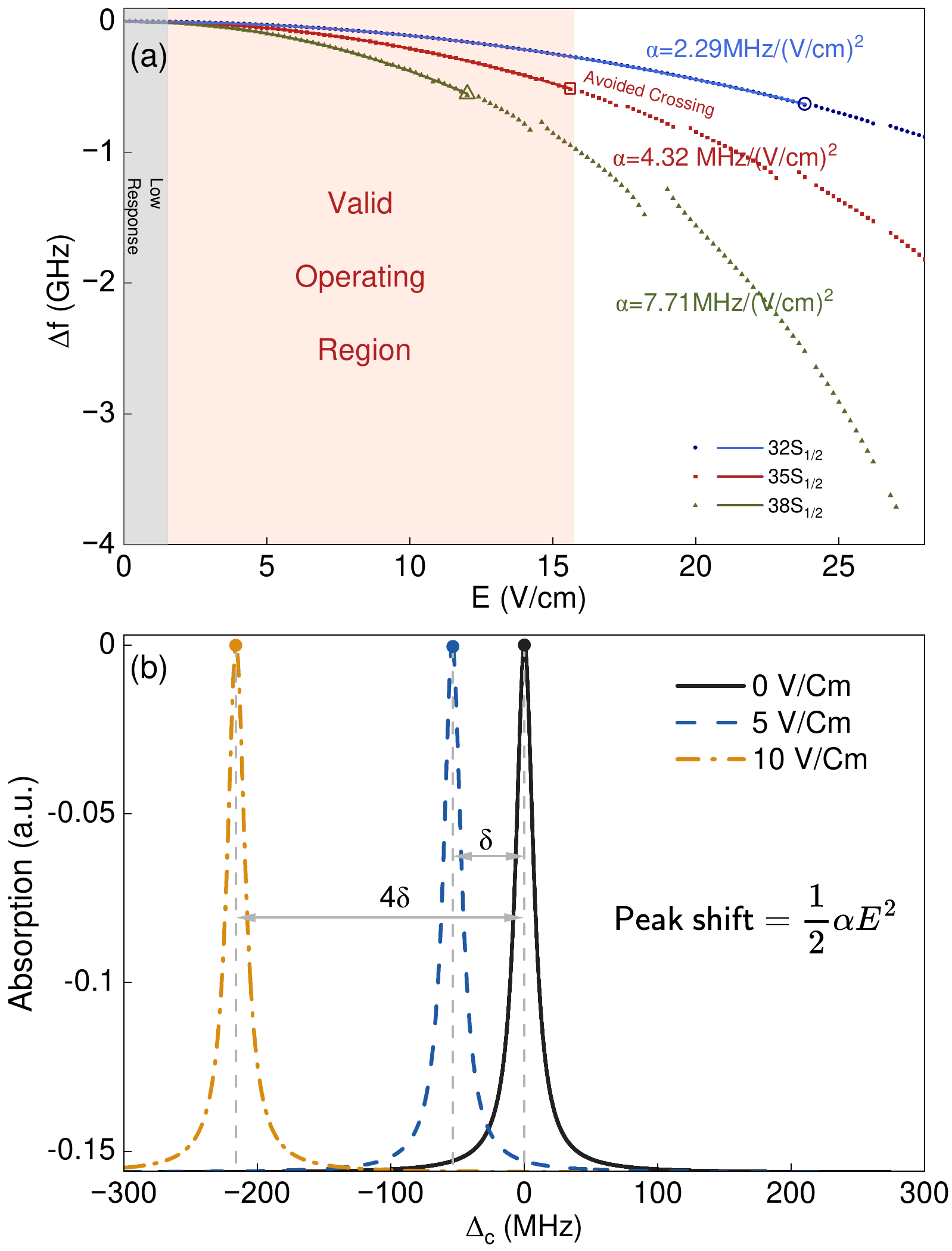}
	\caption{Quadratic Stark response and the corresponding EIT peak shift under DC electric fields. (a) Calculated DC Stark shifts of the Rb $32\mathrm{S}_{1/2}$, $35\mathrm{S}_{1/2}$, and $38\mathrm{S}_{1/2}$ Rydberg states as functions of the electric field, obtained from the open-source ARC Python package~\cite{Sibalic2017}. The grey and red shaded regions denote the weak-field insensitive regime and the practical quadratic operating range, respectively. (b) Simulated EIT absorption spectra versus coupling detuning $\Delta_c$ for different DC field strengths ($0$, $5$, and $10~\mathrm{V/cm}$), showing that the resonance peak shifts with an approximately quadratic dependence on the applied field. The labels $\delta$ and $4\delta$ illustrate the quadratic scaling of the peak displacement with the field amplitude.}
	\label{fig:DC Stark}
\end{figure}

The choice of Rydberg state is therefore a key design consideration for measuring low-frequency electric fields. As shown in Fig.~\ref{fig:DC Stark}(a), the calculated DC Stark shifts of the Rb $32\mathrm{S}_{1/2}$, $35\mathrm{S}_{1/2}$, and $38\mathrm{S}_{1/2}$ states all exhibit an approximately quadratic dependence on the electric field in the low- to intermediate-field regime, while their polarizabilities increase with principal quantum number. This trend is beneficial for sensitivity, because a larger polarizability produces a larger Stark shift under the same applied field and thus yields a stronger spectral response. At the same time, however, higher-lying Rydberg states are also more susceptible to field-induced mixing with nearby levels, so their Stark maps depart from the simple parabolic behavior at comparatively lower fields, leading more readily to crossings or avoided crossings. As a result, although a higher Rydberg state generally provides greater field sensitivity, its usable quadratic operating range becomes narrower. By contrast, lower-lying states usually preserve a broader and more robust quadratic region, but their smaller polarizabilities lead to weaker Stark shifts under the same applied field. The choice of sensing state therefore involves a trade-off between Stark sensitivity and the width of the usable electric-field range. In this work, the Rb $35\mathrm{S}_{1/2}$ state is adopted as the representative sensing level because it provides a suitable compromise between these two requirements. Fitting the quadratic part of the $35\mathrm{S}_{1/2}$ Stark curve yields a polarizability of $\alpha = 4.32~\mathrm{MHz}/(\mathrm{V/cm})^2$, which is used in the subsequent sensing analysis. For this representative state, the corresponding practical operating region is highlighted by the red shaded area in Fig.~\ref{fig:DC Stark}(a).

Once the sensing state is selected, the low-frequency electric field is transduced through the Stark-induced displacement of the EIT resonance rather than through ATS splitting. As shown in Fig.~\ref{fig:DC Stark}(b), when the DC field increases of $0$, $5$, and $10~\mathrm{V/cm}$, the EIT transparency peak shifts progressively away from $\Delta_c=0$, and the resonance displacement follows an approximately quadratic dependence on the applied field. When the Stark-induced EIT peak shift is used as the measurement observable, the electric-field amplitude can in principle be inferred as
\begin{equation}
	E = \sqrt{\frac{2\Delta S}{\alpha}}.
\end{equation}

However, direct peak-shift readout based on this quadratic Stark response is still subject to intrinsic limitations in the weak-field regime. First, the response slope associated with Eq.~(9), $\mathrm{d}\Delta S/\mathrm{d}E=\alpha E$, becomes increasingly small as the electric field decreases and ultimately vanishes in the zero-field limit. As a result, the Stark-induced displacement becomes too small to provide a sufficiently sensitive readout for weak electric fields. This weak-response region is indicated by the grey shaded area in Fig.~\ref{fig:DC Stark}(a). Second, practical peak tracking further requires the Stark-induced frequency shift to be comparable to the EIT FWHM. As a representative estimate, using a typical FWHM of $4.0~\mathrm{MHz}$ together with a polarizability of $4.32~\mathrm{MHz}/(\mathrm{V/cm})^2$ for the Rb $35\mathrm{S}_{1/2}$ state gives a minimum resolvable field of approximately $1.36~\mathrm{V/cm}$. This threshold remains too high for many practical low-frequency electric-field sensing tasks in power-system environments.

\section{FI and CRLB for EIT-Based Electric-Field Sensing}

To quantify the sensitivity and precision limits of Rydberg-atom EIT electric field sensing, we employ the FI and CRLB framework. Section~\ref{subsec:FI_CRLB} summarizes the general definitions of FI and the CRLB, and discusses their physical meanings and mutual connection. Section~\ref{subsec:EIT_FI} then applies this framework to the EIT spectra and derives an explicit FI expression in terms of the probe transmittance.

\subsection{FI and CRLB: Definitions and Interpretation}
\label{subsec:FI_CRLB}

In classical statistics, FI\cite{Ly2017,liu2025sensitivitylimitrydbergelectrometry,Tsai2023} quantifies how sensitively measurement data respond to variations in an unknown parameter. A larger FI implies that the data are more informative about the parameter and thus enable a more precise estimation. FI is a key quantity for characterizing sensitivity and information content in both classical and quantum parameter estimation. It is directly linked to the CRLB, which sets a fundamental lower bound on the variance of any unbiased estimator under suitable regularity conditions. Together, FI and CRLB provide a basic framework for parameter estimation: FI quantifies information in the data, whereas the CRLB specifies the corresponding precision limit. In the present work, this framework is used for operating-point selection and sensitivity evaluation in the EIT sensing scheme.

Let $p(x|\theta)$ denote the probability distribution of an observation $x$ parameterized by an unknown parameter $\theta$. In the continuous case, $p(x|\theta)$ is a probability density function, whereas in the discrete case it is a probability mass function. For a single observation $x$, the log-likelihood is $L(x;\theta)=\ln p(x|\theta)$, which simplifies differentiation without changing the location of the maximum. Its derivative with respect to $\theta$,
\begin{equation}
	S(x;\theta)
	= \frac{\partial}{\partial \theta}\,L(x;\theta)
	= \frac{\partial}{\partial \theta}\ln p(x|\theta),
\end{equation}
is referred to as the score function. Intuitively, $S(x;\theta)$ quantifies how sensitively the log-likelihood of the observed outcome $x$ changes with the parameter $\theta$. Ultimately, the FI is defined as the expectation of the squared score,
\begin{equation}
	F(\theta)
	= \mathbb{E}_\theta\!\left[S(x;\theta)^2\right]
	= \int p(x|\theta)
	\left[
	\frac{\partial}{\partial \theta}
	\ln p(x|\theta)
	\right]^2 \mathrm{d}x,
	\label{Eq:FI_definition}
\end{equation}
which provides a quantitative measure of the average information that the observational data carries about $\theta$; equivalently, it characterizes the expected curvature of the log-likelihood function. Here the expectation $\mathbb{E}_\theta[\cdot]$ is taken with respect to $p(x|\theta)$; in the discrete case, the integral is replaced by a summation over all possible values of $x$.

The connection between FI and estimation accuracy is established via the CRLB. Under appropriate regularity conditions, any unbiased estimator $\hat{\theta}$ of the parameter $\theta$ satisfies\cite{Ly2017}
\begin{equation}
	\mathrm{Var}(\hat{\theta}) \ge \mathrm{CRLB}(\theta) = \frac{1}{F(\theta)},
	\label{Eq:CRLB}
\end{equation}
which is the CRLB for the single-parameter case. This inequality shows that a larger FI leads to a smaller minimum attainable variance, and hence to a higher theoretical measurement sensitivity. For a data set consisting of $N$ independent and identically distributed samples, the total FI typically scales approximately linearly with $N$, so the CRLB yields $\mathrm{Var}(\hat{\theta}) \propto 1/N$. This implies that the measurement uncertainty scales as $1/\sqrt{N}$, which corresponds to the standard quantum limit (SQL).

With the relationship between FI and the CRLB established, we apply the FI framework to Rydberg-atom EIT spectra to quantify how the measurement sensitivity depends on system settings, such as the operating point and detunings, as well as on the external electric field amplitude. This provides a quantitative basis for identifying optimal operating conditions, assessing fundamental sensitivity limits, and guiding experimental design and engineering implementation.

\subsection{FI and CRLB in EIT Sensing}
\label{subsec:EIT_FI}

In an EIT measurement, the quantity of interest $\theta$ (e.g., an external field amplitude or an effective resonance shift) is inferred from the transmitted probe signal through its influence on the EIT spectral response. The experimentally accessible observable is the number of transmitted photons $n$ collected within a given detection window for a probe beam with a fixed incident intensity. Within the FI framework, we therefore take $n$ as the observation variable and treat $\theta$ as the unknown parameter, so that the EIT-based parameter estimation problem fits naturally into the general FI definition in Eq.~\eqref{Eq:FI_definition}.

Because the probe laser can be well approximated by coherent light, the associated photon-counting statistics are Poissonian. Accordingly, the transmitted photon number $n$ can be modeled as a Poisson random variable with mean $\bar n$ and probability mass function
\begin{equation}
	P(n|\bar n)=\frac{\bar n^{\,n}e^{-\bar n}}{n!}.
\end{equation}
In an EIT measurement, the mean transmitted photon number is set by the transmission of the three-level system and depends on the unknown parameter, i.e., $\bar n=\bar n(\theta)$. For a single outcome $n$, the log-likelihood is defined as $L(n;\theta)=\ln P\!\bigl(n|\bar n(\theta)\bigr)$, and the corresponding score function is
\begin{equation}
	S(n;\theta)
	=\frac{\partial}{\partial \theta}L(n;\theta)
	=\frac{n-\bar n(\theta)}{\bar n(\theta)}\,
	\frac{\partial \bar n(\theta)}{\partial \theta},
\end{equation}
which quantifies the sensitivity of the log-likelihood to variations in $\theta$. Substituting $S(n;\theta)$ into the FI definition in Eq.~\eqref{Eq:FI_definition} gives
\begin{equation}
	\begin{aligned}
		F(\theta)
		&=\sum_{n} P\bigl(n|\bar n(\theta)\bigr)
		\left(
		\frac{n-\bar n(\theta)}{\bar n(\theta)}
		\right)^2
		\left[
		\frac{\partial \bar n(\theta)}{\partial \theta}
		\right]^2 \\
		&=\left[
		\frac{1}{\bar n(\theta)}
		\right]^2
		\left[
		\frac{\partial \bar n(\theta)}{\partial \theta}
		\right]^2
		\sum_{n} P\bigl(n|\bar n(\theta)\bigr)\,
		\bigl(n-\bar n(\theta)\bigr)^2 .
	\end{aligned}
\end{equation}
The remaining summation is the variance of the Poisson distribution, $\mathrm{Var}(n)=\bar n(\theta)$, which yields the compact result
\begin{equation}
	F(\theta)
	= \frac{1}{\bar n(\theta)}
	\left[
	\frac{\partial \bar n(\theta)}{\partial \theta}
	\right]^2 .
	\label{Eq:FI2}
\end{equation}

Next, we incorporate the EIT transmission response into the FI expression. Let $\eta(\theta)$ denote the probe intensity transmittance, so that the mean transmitted photon number is $\bar n(\theta)=\bar n_0\,\eta(\theta)$, where $\bar n_0$ is the incident mean photon number. Taking the derivative with respect to $\theta$ gives
\begin{equation}
	\frac{\partial \bar n(\theta)}{\partial \theta}
	=\bar n_0\,\frac{\partial \eta(\theta)}{\partial \theta}
	=\bar n(\theta)\,\frac{\partial \ln \eta(\theta)}{\partial \theta}.
\end{equation}
Substituting this result into Eq.~\eqref{Eq:FI2} yields the analytic FI in EIT model\cite{liu2025sensitivitylimitrydbergelectrometry},
\begin{equation}
	F(\theta)
	=\bar n_0\,\eta(\theta)
	\left[
	\frac{\partial \ln \eta(\theta)}{\partial \theta}
	\right]^2 .
	\label{Eq:FI_EIT}
\end{equation}

This FI expression admits a clear physical interpretation. The operating point that maximizes $F(\theta)$ typically balances two requirements: (i) a sufficiently high transmission, i.e., a large value of $\eta(\theta)$, which ensures a large mean photon number $\bar n(\theta)$ and thereby suppresses shot noise arising from Poissonian statistics; and (ii) a large normalized slope $\left|\partial_{\theta}\ln\eta(\theta)\right|$, ensuring that the transmission responds strongly to small variations of the unknown parameter. In practice, the optimal operating point therefore lies on a steep portion of the EIT line shape while maintaining a relatively high transmission level, enabling the highest estimation accuracy for a given photon number budget.

According to Eq.~\eqref{Eq:CRLB} and under suitable regularity conditions, the variance of any unbiased estimator $\hat{\theta}$ satisfies
\begin{equation}
	\mathrm{Var}(\hat{\theta})
	\ge
	\mathrm{CRLB}(\theta)
	=
	\frac{1}{
		\bar n_0\,\eta(\theta)
		\left[
		\dfrac{\partial \ln \eta(\theta)}{\partial \theta}
		\right]^2
	}.
	\label{Eq:CRLB_theta}
\end{equation}
This bound shows that the CRLB directly sets the minimum attainable estimation variance for the parameter $\theta$. Taken together, the FI and CRLB framework provides a quantitative basis for sensitivity evaluation, optimal operating point selection, and experimental design in EIT-based parameter estimation.

\begin{figure*}
	\centering  
	\includegraphics[width=0.75\linewidth]{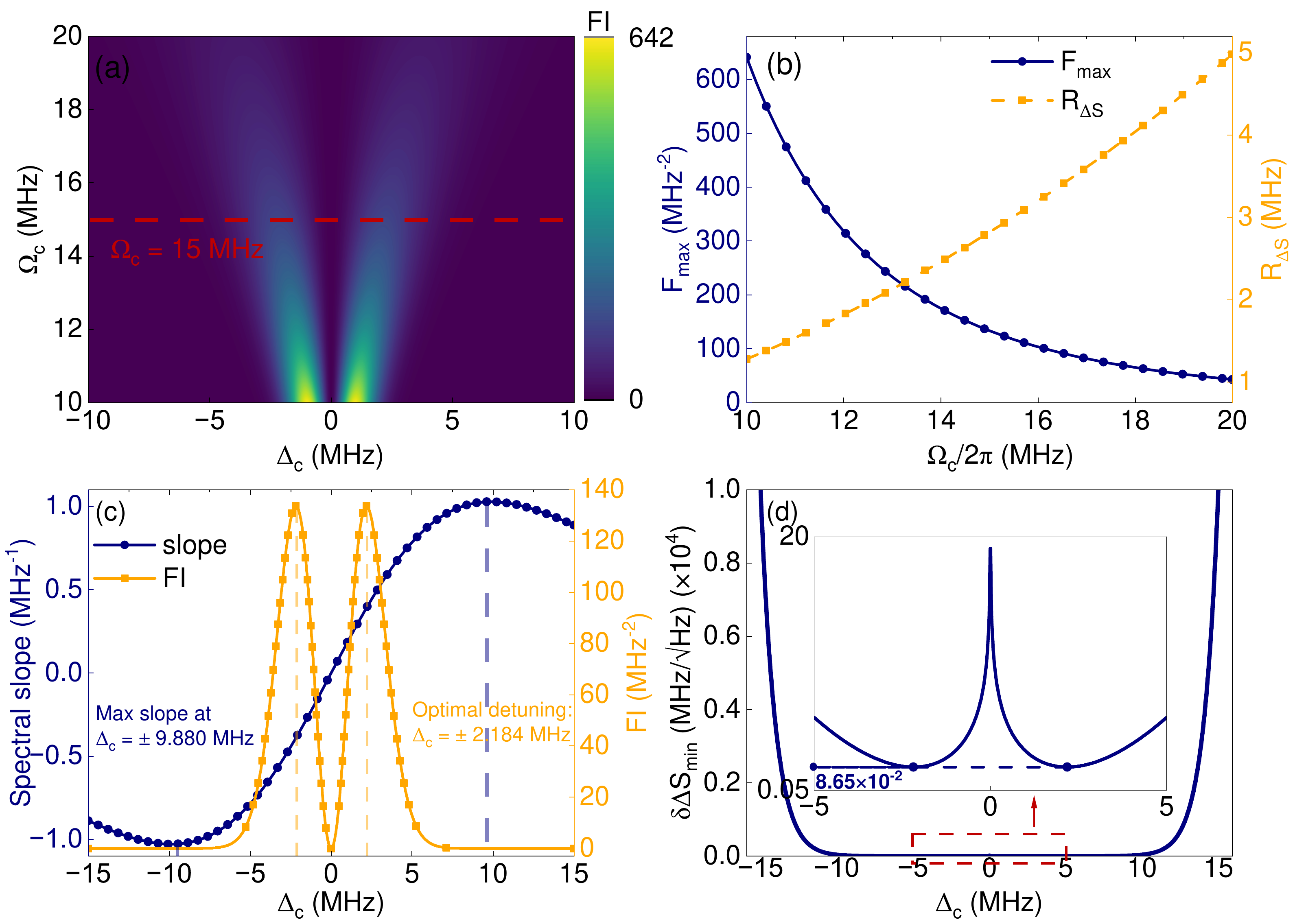}
	\caption{FI analysis of Stark shift estimation in the cascade EIT system. (a) FI map in the $(\Delta_c,\Omega_c)$ plane. The red dashed line marks the representative slice at $\Omega_c/2\pi=15~\mathrm{MHz}$ used in panels (c) and (d). (b) Peak FI $F_{\max}$ and usable Stark-shift range $R_{\Delta S}$ versus $\Omega_c$, both evaluated at the FI-optimal operating point. Their opposite trends indicate a trade-off between sensitivity and dynamic range across different coupling strengths. (c) FI and corresponding spectral slope versus $\Delta_c$ at $\Omega_c/2\pi=15~\mathrm{MHz}$. The FI-optimal operating points are located at $\Delta_c=\pm 2.184~\mathrm{MHz}$, whereas the maximum-slope points occur at $\Delta_c=\pm 9.880~\mathrm{MHz}$, showing that the FI optimum does not coincide with the maximum-slope condition. (d) CRLB-limited minimum detectable Stark shift $\delta\Delta S_{\min}$. The inset enlarges the near-resonant region $\Delta_c\in[-5,5]~\mathrm{MHz}$, where the minimum reaches $8.65\times10^{-2}~\mathrm{MHz}/\sqrt{\mathrm{Hz}}$. Parameters: $\Omega_p/2\pi=2~\mathrm{MHz}$, $\Omega_c/2\pi=15~\mathrm{MHz}$, and $\bar n_0\approx 4.7\times10^{14}$.}
	\label{fig:FI_stark}
\end{figure*}

To avoid the weak-field degeneracy introduced by the quadratic Stark relation, we choose the Stark-induced shift $\Delta S$ as the parameter to be estimated, i.e., $\theta = \Delta S$. The coupling detuning $\Delta_c$ is scanned as a controllable operating-point variable to locate the maximum FI. The numerical results are summarized in Fig.~\ref{fig:FI_stark}; unless otherwise specified, all other parameters are the same as in the previous sections.

Figure~\ref{fig:FI_stark}(a) presents the FI distribution in the two-dimensional parameter space spanned by the coupling detuning $\Delta_c$ and the coupling Rabi frequency $\Omega_c$. The map shows that, within the scanned parameter range, the FI increases as $\Omega_c$ decreases, with the high-FI region concentrated toward the lower-$\Omega_c$ side. To further clarify this trend, we extract from Fig.~\ref{fig:FI_stark}(a) the peak FI at each $\Omega_c$, denoted $F_{\max}(\Omega_c)$. As shown in Fig.~\ref{fig:FI_stark}(b), $F_{\max}$ increases monotonically as $\Omega_c$ decreases. This reflects the fact that a smaller coupling Rabi frequency narrows the EIT window, thereby steepening the spectral slope and enhancing the estimation precision. However, a narrower EIT window also reduces the Stark-shift range over which the spectral response remains in its high-sensitivity region. To quantify this effect, we define $R_{\Delta S}(\Omega_c)$ as the usable Stark-shift range at each $\Omega_c$ (see Appendix~\ref{app:dynamic_range} for the formal definition). As plotted in Fig.~\ref{fig:FI_stark}(b), $F_{\max}$ and $R_{\Delta S}$ exhibit opposite trends: decreasing $\Omega_c$ improves the estimation precision but simultaneously narrows the usable operating range. This reveals a fundamental sensitivity--dynamic-range trade-off governed by the coupling strength.

To examine the detuning dependence at a representative coupling strength, we select the slice at $\Omega_c/2\pi=15~\mathrm{MHz}$, as marked by the red dashed line in Fig.~\ref{fig:FI_stark}(a). Figure~\ref{fig:FI_stark}(c) shows the FI and the corresponding spectral slope as functions of $\Delta_c$ along this slice. Two symmetric FI maxima are located at $\Delta_c=\pm 2.184~\mathrm{MHz}$, which define the optimal operating points for Stark-shift estimation, whereas the maximum-slope points occur farther away at $\Delta_c=\pm 9.880~\mathrm{MHz}$. The FI maxima lie closer to the center of the EIT transmission peak than the slope maxima, indicating that the optimal operating point is determined by the trade-off between parametric sensitivity and photon budget rather than by slope alone. This is most evident at $\Delta_c=0$, where the spectral slope vanishes by symmetry, yielding zero FI despite the maximum spectral transmittance. Figure~\ref{fig:FI_stark}(d) plots the resulting CRLB-limited minimum detectable Stark shift $\delta \Delta S_{\min}$ versus $\Delta_c$. In the far-detuned region, the FI is strongly suppressed and the corresponding uncertainty rises rapidly. Focusing on the near-resonant window $\Delta_c\in[-5,5]~\mathrm{MHz}$, as highlighted in the inset, the minimum sensitivity bound is found to be $8.65\times10^{-2}~\mathrm{MHz}/\sqrt{\mathrm{Hz}}$, which sets the theoretical precision limit for Stark-shift estimation under the assumed photon-number budget.

\section{Two-Point Differential EIT Scheme for Electric-Field Measurement}

Having analyzed the estimation precision for the Stark-induced shift $\Delta S$, we now turn to the ultimate quantity of interest in power-system sensing, namely the electric field amplitude $E$. To overcome the poor weak-field sensitivity and the resolution limits of peak-shift readout under a purely quadratic Stark response, we employ a DC-biased two-point differential EIT scheme for electric field measurement. By applying a controllable DC bias and sampling the EIT spectrum at two symmetric detuning points, this approach converts the quadratic Stark dependence into an effectively linear response to the signal field, thereby overcoming the peak-shift resolution limit and substantially enhancing weak-field sensitivity.

\subsection{DC-Field Measurement}
\label{subsec:two_point_DC_en}

\begin{figure}
	\centering  
	\includegraphics[width=\linewidth]{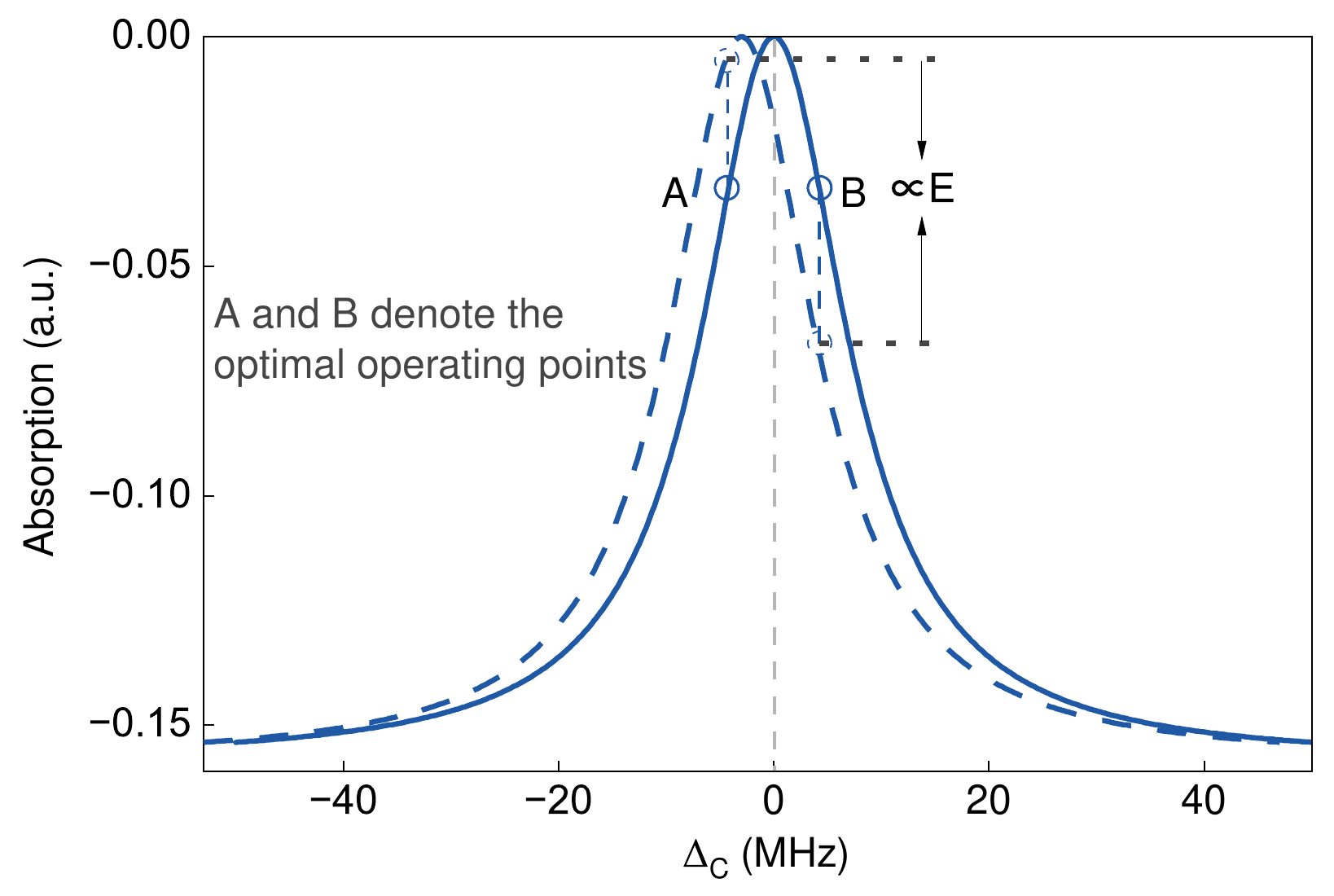}
	\caption{Schematic of the DC-biased two-point differential readout scheme. The solid and dashed blue curves represent the unperturbed and Stark shifted EIT spectra, respectively. Points A and B denote the two symmetric positions with maximum FI, i.e., the optimal operating points.}
	\label{fig:differential}
\end{figure}

We begin with the DC field case. Let $\rho(\Delta_c,E)\equiv \mathrm{Im}\{\rho_{ge}(\Delta_c,E)\}$ denote the absorption signal extracted from the probe coherence, where $E$ is the unknown DC electric field to be measured. In the absence of an external field, the EIT line shape is approximately symmetric about $\Delta_c=0$. As conceptually illustrated in Fig.~\ref{fig:differential}, we select two operating points, $\mathrm{A}$ and $\mathrm{B}$, at detunings $+\delta$ and $-\delta$ (in our scheme, $\delta = 2.184~\mathrm{MHz}$), with the corresponding zero-field baseline signals defined as
\begin{equation}
	\rho_{\mathrm{A}} \equiv \rho(+\delta,0),\qquad
	\rho_{\mathrm{B}} \equiv \rho(-\delta,0),
\end{equation}
so that ideally $\rho_{\mathrm{A}}=\rho_{\mathrm{B}}$. We define the two-point differential signal $\rho_{\mathrm{AB}} \equiv \rho_{\mathrm{A}}-\rho_{\mathrm{B}}$, which vanishes in the field-free case. Beyond being a mathematical convenience, this two-point differential architecture fundamentally acts as a balanced detection mechanism to suppress common-mode noise. In practical EIT setups, low-frequency technical fluctuations, including laser intensity variations, slow frequency drifts, and atomic density changes, affect both symmetric operating points ($+\delta$ and $-\delta$) equally. By evaluating the difference $\rho_{\mathrm{AB}}$, these correlated noise sources are effectively canceled out, preserving the Stark-induced signal.

For a weak applied DC field, the EIT resonance undergoes an effective horizontal shift along the coupling detuning axis, with the shift magnitude set by Eq.~\eqref{Eq:Stark}. In the small-shift limit ($\Delta S\ll\delta$), the field-perturbed spectrum can be approximated as a rigid translation of the field-free line shape, $\rho(\Delta_c,E) \approx \rho_0(\Delta_c-\Delta S)$, where $\rho_0(\Delta_c)\equiv \rho(\Delta_c,0)$. Substituting into $\rho_{\mathrm{AB}}$ and performing a Taylor expansion with respect to $\Delta S$, the approximate symmetry of the unperturbed line shape about $\Delta_c=0$ causes the constant baselines and all even-order terms to cancel. The resulting differential signal is dominated by the first-order contribution, yielding
\begin{equation}
	\rho_{\mathrm{AB}}(E)\approx -2\,\rho_0'(\delta)\,\Delta S = -\,\alpha\,\rho_0'(\delta)\,E^2.
	\label{eq:TPD}
\end{equation}
The DC-field amplitude can then be inferred as
\begin{equation}
	E=\sqrt{\frac{|\rho_{\mathrm{AB}}(E)|}{\alpha\,|\rho_0'(\delta)|}},
\end{equation}
where the absolute values are taken since the measurement depends only on the field magnitude.

Although the two-point differential scheme enables a robust extraction of the EIT response in the weak-field regime, the resulting signal remains intrinsically quadratic in $E$ due to the Stark relation. To obtain an approximately linear readout for small fields, we superimpose a known DC bias and use a differential measurement around the bias point.

Specifically, we apply two DC bias fields of equal magnitude and opposite polarity, $+E_0$ and $-E_0$, such that the total fields are $E_{+}=E_0+E$ and $E_{-}=-E_0+E$, respectively. Evaluating the two-point differential signal under each bias polarity via Eq.~\eqref{eq:TPD} gives $\rho_{\mathrm{AB}}(E_{+})$ and $\rho_{\mathrm{AB}}(E_{-})$. Forming the bias-reversal differential signal $\rho_{\mathrm{AB}}(E_\Delta)\equiv \rho_{\mathrm{AB}}(E_{+})-\rho_{\mathrm{AB}}(E_{-})$ then eliminates the quadratic term and yields an effectively linear response,
\begin{equation}
	\rho_{\mathrm{AB}}(E_\Delta)\approx -4\,\alpha\,\rho_0'(\delta)\,E_0\,E .
\end{equation}
Accordingly, the unknown field magnitude can be retrieved via
\begin{equation}
	|E|\approx \frac{\bigl|\rho_{\mathrm{AB}}(E_\Delta)\bigr|}{4\,\alpha\,\bigl|\rho_0'(\delta)\bigr|\,|E_0|}.
\end{equation}

In this way, bias reversal combined with two-point differential sampling effectively linearizes the originally quadratic Stark response. Overall, the procedure is as follows: the FI analysis is first used to determine the optimal symmetric operating points $\pm\delta$; the two-point differential signal $\rho_{\mathrm{AB}}$ is then evaluated under the two bias configurations $E_{+}$ and $E_{-}$; finally, the bias-reversal differencing $\rho_{\mathrm{AB}}(E_{\Delta})$ yields a readout that is linear in $E$ in the weak-field regime. This linearized readout thus provides a practical pathway for weak-field sensing in low-frequency electric field measurements.

\begin{figure}
	\centering  
	\includegraphics[width=\linewidth]{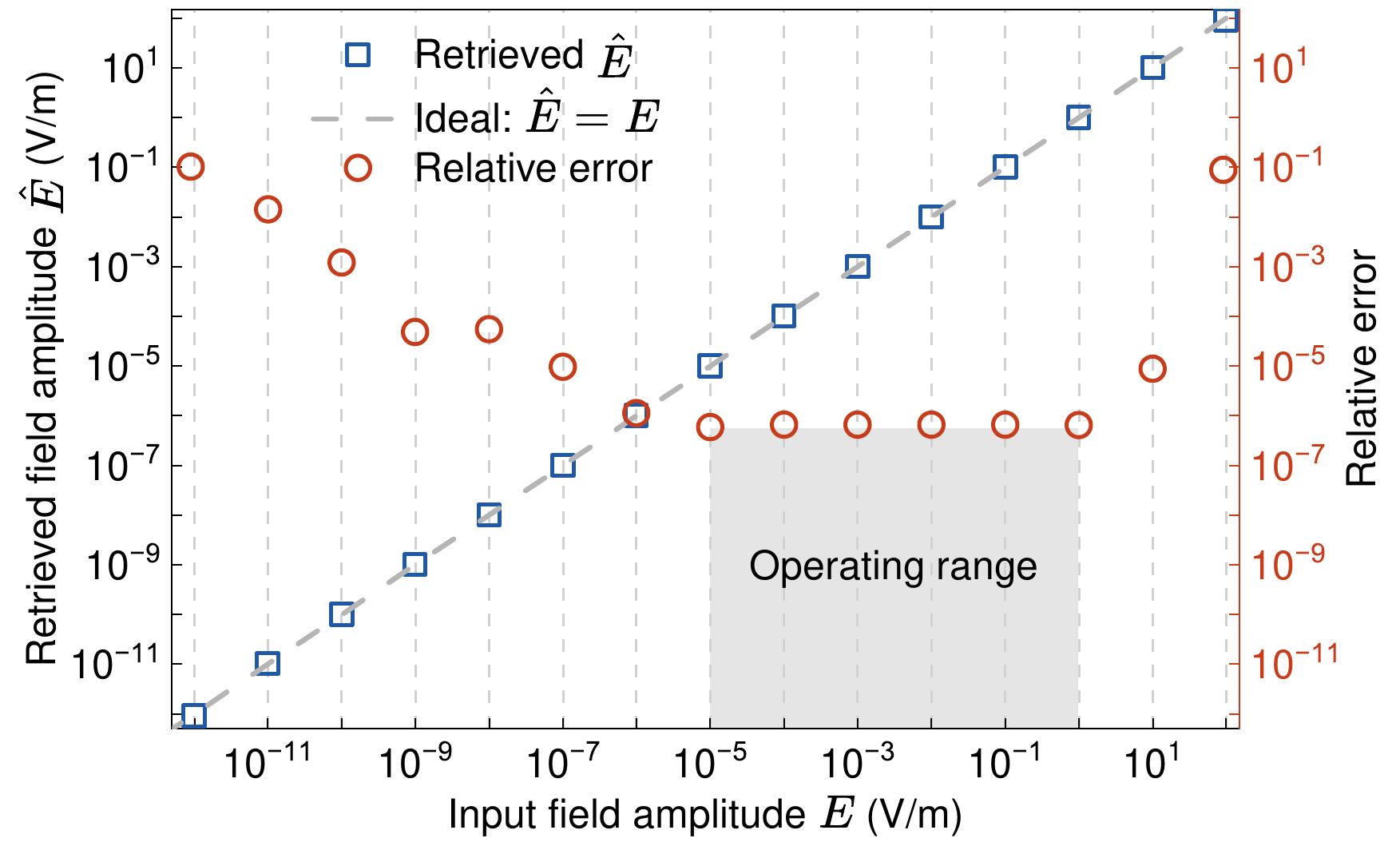}
	\caption{Numerical validation of the DC-biased two-point differential readout. The prescribed DC field amplitude $E$ (horizontal axis) is compared with the retrieved value $\hat{E}$ (open blue squares) obtained from the simulated measurement procedure; the grey dashed line indicates the ideal relation $\hat{E}=E$. The open red circles, corresponding to the right vertical axis, show the relative error. The grey shaded region highlights the sensor's operating range. Both axes use logarithmic scales to cover the wide dynamic range. Simulation parameters include a DC bias field $E_0=1~\mathrm{V/m}$.}
	\label{fig:retrieve_E}
\end{figure}

Figure~\ref{fig:retrieve_E} summarizes a numerical validation of the proposed measurement procedure. Because the input field amplitude spans many orders of magnitude, both axes are plotted on logarithmic scales. The horizontal axis shows the prescribed reference field amplitude $E$, while the open blue squares report the retrieved value $\hat{E}$ obtained from the simulated DC-biased two-point differential readout. As seen in the figure, $\hat{E}$ agrees exceptionally well with $E$ over a wide dynamic range, closely following the ideal relation $\hat{E}=E$ (indicated by the grey dashed line). Furthermore, to quantitatively assess the retrieval accuracy, we plot the corresponding relative error (open red circles, right axis). The grey shaded area highlights the optimal operating range of the sensor (approximately from $10^{-5}$ to $1~\mathrm{V/m}$), where the relative error remains as low as $10^{-7}$. Beyond this range, the error increases either due to numerical resolution limits at very low fields or the breakdown of the small-shift approximation at high fields.

In these simulations, we choose a DC bias field of $E_0=1~\mathrm{V/m}$. The corresponding Stark-induced shift is $\Delta S\simeq 216~\mathrm{Hz}$, which is far below the MHz-scale spectral features of the EIT line, and is therefore consistent with our assumption. It should be emphasized that the ultra-low electric field values appearing in Fig.~\ref{fig:retrieve_E} do not represent the actual detection limit of the sensor. Figure~\ref{fig:retrieve_E} only provides a numerical validation of the proposed retrieval procedure under ideal noise-free conditions, and therefore demonstrates the self-consistency of the retrieval model in the weak-field regime. By contrast, the minimum resolvable electric field must be evaluated within a statistical estimation framework that explicitly accounts for the finite photon number and the photon-counting statistics.

\begin{figure}
	\centering  
	\includegraphics[width=\linewidth]{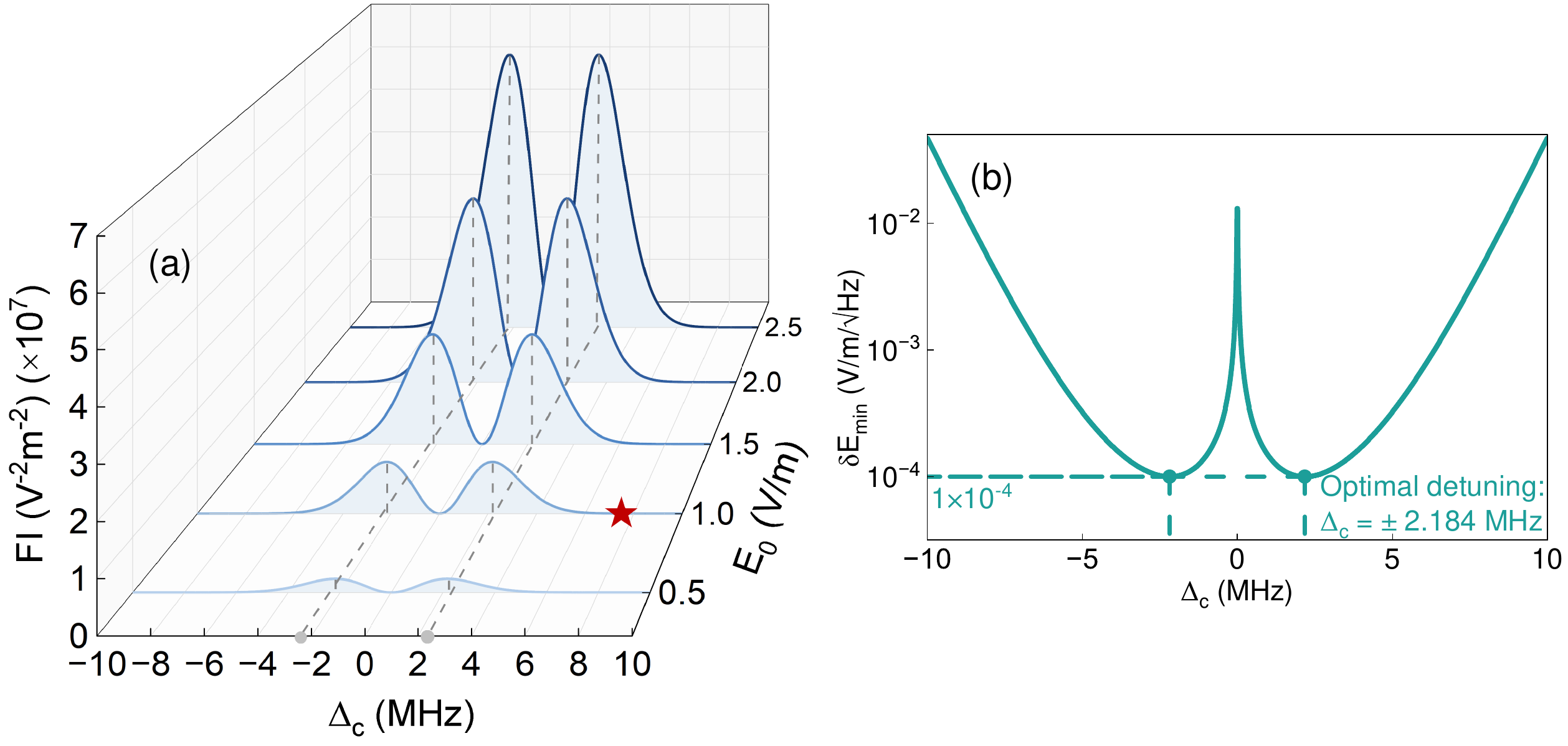}
	\caption{FI optimization and CRLB-limited resolution for the DC-biased two-point differential readout. (a) Waterfall plot of the FI versus coupling detuning $\Delta_c$ for different DC bias fields $E_0$. The dashed guide lines indicate the nearly fixed optimal detunings, and the red pentagram marks the bias value $E_0=1~\mathrm{V/m}$ selected for the subsequent analysis. (b) Corresponding CRLB-limited minimum detectable electric field $\delta E_{\min}$ versus $\Delta_c$ at the selected bias. Two symmetric minima occur at $\Delta_c=\pm 2.184~\mathrm{MHz}$, and the minimum theoretical sensitivity bound reaches approximately $1\times10^{-4}~\mathrm{V/m}/\sqrt{\text{Hz}}$.}
	\label{fig:DC_FI}
\end{figure}

To this end, we next quantify the achievable field resolution of the DC-biased two-point differential scheme within the FI and CRLB framework by taking the unknown DC field amplitude as the parameter of interest, i.e., $\theta=E$. Under this model, the FI can be written as
\begin{equation}
	F(E)
	=\bar n_0\,\eta(E)
	\left[
	\frac{\partial \ln \eta(E)}{\partial \Delta S}\,
	\frac{\partial \Delta S}{\partial E}
	\right]^2 ,
	\label{Eq:FI_chain_DS}
\end{equation}
where the derivative with respect to $E$ is evaluated via the chain rule. Substituting $\eta(E)$ using Eqs.~\eqref{chi} and \eqref{eta}, and using the weak-field expansion about the bias point so that $\partial \Delta S/\partial E=\alpha E_0$. Moreover, the four measurements at operating points $\mathrm{A}$ and $\mathrm{B}$ under each of the two bias polarities are performed with independent photon-counting windows, so their FI contributions are additive and the total FI acquires an overall factor of $4$, yielding
\begin{equation}
	\begin{split}	
	F(E)&=\,4\,\bar{n}_0\,\exp \!\left[ \beta \,\rho \left( \delta \right) \right]\\
	&\times \left( \beta \,\alpha E_0 \right) ^2\left[ \frac{\partial \rho \left( \delta \right)}{\partial \Delta S} \right] ^2,\\	
	\end{split}
	\label{Eq:FI_bias_final}
\end{equation}

where $\beta \equiv \frac{2N k_p l \mu_{ge}^2}{\varepsilon_0 \hbar \Omega_p}$
collects the constant prefactors of the optical response. Since the CRLB bounds the estimation variance, the minimum resolvable field amplitude is obtained as $\delta E_{\min}=1/\sqrt{F(E)}$, giving
\begin{equation}
\delta E_{\min}=\frac{e^{-\frac{1}{2}\beta \rho \left( \delta \right)}}{2\sqrt{\bar{n}_0}\,\beta \,\alpha E_0}\left| \frac{\partial \,\rho \left( \delta \right)}{\partial \Delta S} \right|^{-1}.
\end{equation}
The numerically evaluated FI and CRLB results are summarized in Fig.~\ref{fig:DC_FI}. Figure~\ref{fig:DC_FI}(a) shows the FI as a function of the coupling detuning $\Delta_c$ for different DC bias fields $E_0$. As the bias increases, the FI profile is progressively enhanced, while the optimal detunings remain nearly unchanged, as indicated by the dashed guide lines. Consistent with the bias value adopted in the preceding numerical validation, we select $E_0=1~\mathrm{V/m}$ for the subsequent analysis, as marked by the red pentagram in Fig.~\ref{fig:DC_FI}(a). The corresponding $\delta E_{\min}$ is plotted in Fig.~\ref{fig:DC_FI}(b). Two symmetric minima occur at $\Delta_c=\pm 2.184~\mathrm{MHz}$, identifying the optimal operating points for the DC-biased two-point differential readout. In contrast, around the EIT center ($\Delta_c\approx 0$), the signal slope becomes vanishingly small, leading to a suppressed FI and a significantly enlarged measurement uncertainty. Under the chosen parameters, the minimum theoretical sensitivity bound reaches approximately $\delta E_{\min}\approx 1\times10^{-4}~\mathrm{V/m}/\sqrt{\text{Hz}}$.

\subsection{AC-Field Measurement}
\label{subsec:AC_field}

After establishing a quantitative relationship between the DC electric field amplitude and the Stark-induced shift of the EIT spectrum, we further investigate the interaction between AC electric fields and the EIT response. Under an AC field modulation, the EIT resonance exhibits a time-dependent periodic frequency oscillation. To enable precise retrieval of the field parameters, we lock the coupling laser detuning to selected operating points ($\mathrm{A}$ and $\mathrm{B}$) on the EIT line shape. In this configuration, the probe beam transmission is converted into a time-varying periodic signal. By demodulating and processing the differential signal between these two points, the frequency and amplitude of the AC electric field can be inferred.

We first consider a purely AC electric field,
\begin{equation}
	E(t)=A\cos(2\pi f_{\mathrm{ac}} t+\phi),
\end{equation}
where $A$ is the field amplitude, $f_{\mathrm{ac}}$ is the field frequency, and $\phi$ is the initial phase. Substituting this into Eq.~\eqref{eq:TPD} yields
\begin{equation}
	\begin{aligned}
		\rho_{\mathrm{AB}}(t)
		&\approx -\,\frac{\alpha\,\rho_0'(\delta)\,A^2}{2}
		\Bigl[1+\cos\!\bigl(4\pi f_{\mathrm{ac}} t+2\phi\bigr)\Bigr].
	\end{aligned}
\end{equation}
It is worth noting that substituting $E(t)$ directly into the steady-state response derived in Eq.~\eqref{eq:TPD} relies on the adiabatic approximation. The dynamics of the Rydberg EIT system to reach a steady state are governed by the atomic relaxation rates (e.g., the intermediate state decay rate $\gamma_e$ and transit-time broadening, typically on the order of MHz), meaning the system equilibrates within a microsecond timescale. In contrast, the period of the targeted power-frequency AC fields (e.g., 50 Hz or 60 Hz) is in the tens of milliseconds range, which is orders of magnitude slower. Because the external field varies so gradually relative to the rapid atomic response, the system adiabatically follows the instantaneous field amplitude. This large separation of timescales rigorously justifies the use of the steady-state susceptibility solution for tracking dynamic low-frequency electric fields.

Applying the Fast Fourier Transform (FFT) to $\rho_{\mathrm{AB}}(t)$ yields
\begin{equation}
	\begin{aligned}
		\rho_{\mathrm{AB}}(f) 
		&= \frac{\alpha\,\rho_0'(\delta)\,A^2}{2}\,\delta(f)
		+ \frac{\alpha\,\rho_0'(\delta)\,A^2}{4}\,e^{i2\phi}\,\delta\!\bigl(f-2f_{\mathrm{ac}}\bigr) \\
		&\quad
		+ \frac{\alpha\,\rho_0'(\delta)\,A^2}{4}\,e^{-i2\phi}\,\delta\!\bigl(f+2f_{\mathrm{ac}}\bigr).
	\end{aligned}
\end{equation}
The spectrum thus contains a DC component and a second-harmonic term at $2f_{\mathrm{ac}}$.
As a consequence, the fundamental frequency $f_{\mathrm{ac}}$ does not appear directly in the readout, which may lead to ambiguity in frequency identification; moreover, the amplitude inference remains limited by the underlying quadratic response.

\begin{figure*}
	\centering
	\includegraphics[width=0.75\linewidth]{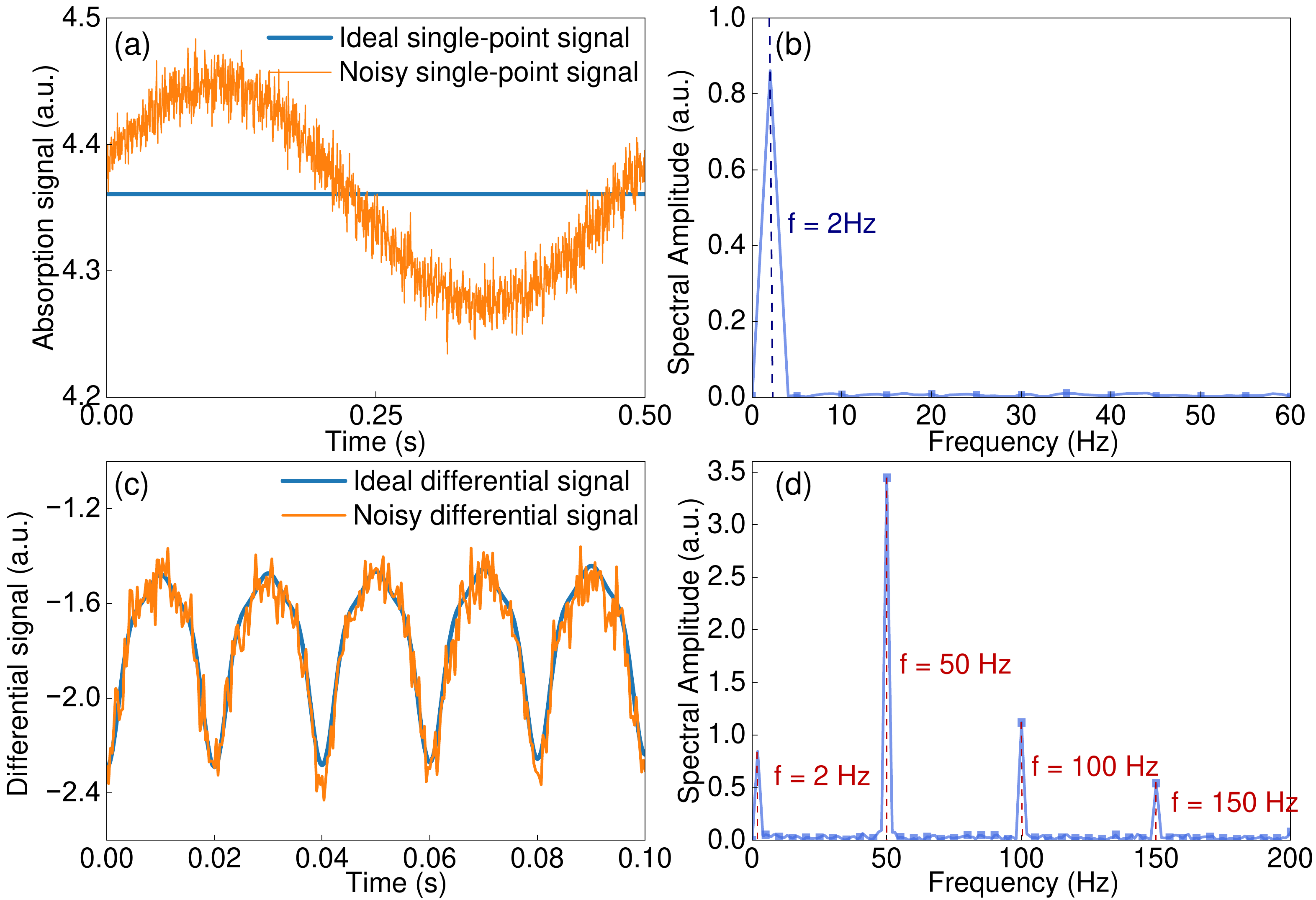}
	\caption{Noise robustness comparison between the baseline and proposed readout schemes. (a) Time-domain single-point signal without DC bias; the weak AC variation is buried in the static background and noise. (b) Corresponding FFT spectrum, where only the noise floor is visible and no distinct peak appears at the target frequency $f_{\mathrm{ac}}=50~\mathrm{Hz}$. (c) Time-domain signal of the proposed scheme, where the target oscillation remains clearly visible under identical noisy conditions. (d) Corresponding FFT spectrum, showing a dominant peak at $f_{\mathrm{ac}}=50~\mathrm{Hz}$; all prescribed harmonic and low-frequency background components are clearly identifiable. Simulation parameters: $E_0=1~\mathrm{V/m}$, $A=0.1~\mathrm{V/m}$, $A_2=0.03~\mathrm{V/m}$, $A_3=0.015~\mathrm{V/m}$, $A_d=0.02~\mathrm{V/m}$, $f_d=2~\mathrm{Hz}$; multiplicative noise: $m_I=0.02$, $f_I=2~\mathrm{Hz}$, $\sigma_I=0.003$; additive detector noise $n_{\mathrm{A,B}}$ root mean square set to $40\%$ of the ideal signal amplitude.}
	\label{fig:AC_t2f}
\end{figure*}

Similarly, to overcome the above limitations, we apply a DC bias $E_0$ and write the total field as
\begin{equation}
	E(t)=E_0 + A\cos\!\bigl(2\pi f_{\mathrm{ac}} t+\phi\bigr). \label{eq:AC_field}
\end{equation}
Substituting the expression into Eq.~\eqref{eq:TPD} yields
\begin{equation}
	\begin{aligned}
		\rho _{\mathrm{AB}}(t) \approx &-\alpha \rho _0' (\delta )\!\left( E_{0}^{2}+\frac{A^2}{2} \right) \\
		&- 2\alpha \rho _0' (\delta )\,E_0A\cos \!\bigl( 2\pi f_{\mathrm{ac}}t+\phi \bigr) \\
		&- \frac{\alpha \rho _0' (\delta )\,A^2}{2}\cos \!\bigl( 4\pi f_{\mathrm{ac}}t+2\phi \bigr).
	\end{aligned}
	\label{eq:expansion}
\end{equation}
The FFT spectrum of $\rho_{\mathrm{AB}}(t)$ is given by
\begin{equation}
	\begin{aligned}
		\rho_{\mathrm{AB}}(f) &= -\,\alpha\,\rho_0'(\delta) \biggl\{ \!\left(E_0^2+\frac{A^2}{2}\right)\delta(f) \\
		&\quad + E_0A \Bigl[e^{i\phi}\delta\!\bigl(f-f_{\mathrm{ac}}\bigr) + e^{-i\phi}\delta\!\bigl(f+f_{\mathrm{ac}}\bigr)\Bigr] \\
		&\quad + \frac{A^2}{4} \Bigl[e^{i2\phi}\delta\!\bigl(f-2f_{\mathrm{ac}}\bigr) + e^{-i2\phi}\delta\!\bigl(f+2f_{\mathrm{ac}}\bigr)\Bigr] \biggr\}.
	\end{aligned}
\end{equation}
From the first-harmonic coefficient at $f_{\mathrm{ac}}$, the field amplitude can be extracted via
\begin{equation}
	A
	= \frac{\bigl|\rho_{\mathrm{AB}}(f_{\mathrm{ac}})\bigr|}
	{\alpha\,\bigl|\rho_0'(\delta)\bigr|\,E_0}.
	\label{eq:amplitude_retrieval}
\end{equation}
Thus, the introduced DC bias effectively linearizes the quadratic response, manifesting as a dominant first-harmonic component. This enables the linear detection of weak low-frequency AC fields utilizing the established two-point differential EIT scheme.

To further evaluate the practicality of the proposed readout under non-ideal conditions, we incorporate both measurement-chain noise and external-field background perturbations into the numerical simulations, so as to better emulate the experimental readout process in a realistic power-system environment. The noisy outputs at the two operating points are modeled as
\begin{equation}
	\tilde{\rho}_{\mathrm{A,B}}(t)=\bigl[1+\epsilon_I(t)\bigr]\,\rho_{\mathrm{A,B}}(t)+n_{\mathrm{A,B}}(t),
\end{equation}
where $\rho_{\mathrm{A,B}}(t)$ is the noise-free readout defined previously and the tilde denotes the noisy counterpart. The measured differential signal is then $\tilde{\rho}_{\mathrm{AB}}(t)=\tilde{\rho}_{\mathrm{A}}(t)-\tilde{\rho}_{\mathrm{B}}(t)$. The multiplicative term $\epsilon_I(t)$ represents probe-intensity fluctuation and laser relative intensity noise associated with the shared optical source and common optical path, and is therefore treated as a common-mode perturbation acting simultaneously on the two channels. In the simulations, it is described by a slowly varying drift term together with a random component,
\begin{equation}
	\epsilon_I(t)=m_I\sin(2\pi f_I t+\varphi_I)+\sigma_I\,\xi_I(t),
\end{equation}
where $m_I$, $f_I$, and $\varphi_I$ denote the modulation depth, characteristic fluctuation frequency, and initial phase, respectively, $\sigma_I$ is the standard deviation of the random component, and $\xi_I(t)$ is a zero-mean unit Gaussian random process. By contrast, the additive terms $n_{\mathrm{A,B}}(t)$ model the independent detector and electronic noise of each readout channel, including photodetector, amplifier, and acquisition noise, and are taken as independent zero-mean Gaussian processes with equal variance. The noise amplitudes are chosen on a relative scale with respect to the characteristic amplitude of the ideal readout signal, so as to assess the robustness of the proposed scheme rather than to reproduce the exact noise level of a specific experimental setup.

In addition, to emulate the external-field complexity in a practical power-system scenario, the target AC field is superimposed with harmonic and low-frequency background components before the quadratic Stark mapping. Specifically, besides the target power-frequency component at $f_{\mathrm{ac}}$, the total electric field includes its second and third harmonics as well as a slowly varying low-frequency perturbation,
\begin{equation}
	\begin{aligned}
		E_{\mathrm{tot}}(t)& = \, A\cos(2\pi f_{\mathrm{ac}} t+\phi) \\
		&+ \sum_{k=2}^{3}A_k\cos(2\pi k f_{\mathrm{ac}}\,t+\phi_k)\\
		&+ A_d\cos(2\pi f_d t+\phi_d),
	\end{aligned}	
\end{equation}
where $A_k$ and $\phi_k$ are the amplitude and phase of the $k$-th harmonic, and $A_d$ and $f_d$ characterize the low-frequency background perturbation. This treatment enables the numerical simulations to simultaneously account for both technical detection noise and realistic environmental field contamination.

The corresponding numerical results are shown in Fig.~\ref{fig:AC_t2f}. In the baseline configuration without DC bias and without differential detection, the single-point signal is dominated by the static background and measurement noise, so that the weak field-induced variation is hardly distinguishable in the time domain, as shown in Fig.~\ref{fig:AC_t2f}(a). The corresponding FFT spectrum in Fig.~\ref{fig:AC_t2f}(b) likewise shows that no distinct peak can be identified at the target frequency $f_{\mathrm{ac}}=50~\mathrm{Hz}$; only the noise floor is visible across the spectrum. This behavior is consistent with the quadratic-response mechanism discussed above: in the weak-field regime, the unbiased single-point readout remains insufficiently sensitive to the target field.

By contrast, once the DC bias is introduced and the two-point differential readout is applied, the noise is largely suppressed by the differential operation. As a result, the target oscillation remains clearly visible in the time domain, as shown in Fig.~\ref{fig:AC_t2f}(c). More importantly, the FFT spectrum in Fig.~\ref{fig:AC_t2f}(d) exhibits a pronounced peak at the fundamental frequency $f_{\mathrm{ac}}=50~\mathrm{Hz}$, together with a weaker component at $2f_{\mathrm{ac}}=100~\mathrm{Hz}$. These results confirm that the bias-assisted differential scheme preserves the linearized first-harmonic response predicted by Eq.~\eqref{eq:expansion}, and significantly improves the visibility and spectral resolvability of weak low-frequency AC electric fields under noisy conditions.

Analogous to the DC case, we quantify the achievable field resolution of the proposed AC sensing scheme within the same FI framework. Here, we take the unknown AC field amplitude as the parameter of interest, i.e., $\theta = A$. Given the time-dependent response derived in \eqref{eq:AC_field}, the FI accumulated over a measurement time $T$ can be expressed as
\begin{equation}
	\begin{aligned}
		F(A) &= \int_0^T 2 \bar{n}_0 \exp \!\left[ \beta \rho(\delta) \right] \\
		&\quad \times \left[ \beta \alpha E_0 \cos \bigl( 2\pi f_{\mathrm{ac}}t + \phi \bigr) \right]^2 \left[ \frac{\partial \rho(\delta)}{\partial \Delta S} \right]^2 dt \\
		&= n_{\text{total}} \exp \!\left[ \beta \rho(\delta) \right] \left[ \beta \alpha E_0 \frac{\partial \rho(\delta)}{\partial \Delta S} \right]^2,
	\end{aligned}
	\label{eq:FI_calculation}
\end{equation}
where $n_{\text{total}}=\bar{n}_0 T$ denotes the total number of photons accumulated over the measurement duration $T$. Equation~\eqref{eq:FI_calculation} shows that the FI for the AC field amplitude is obtained by integrating the instantaneous contribution over the observation window. The prefactor of $2$ reflects the independent contributions from the two symmetric operating points $\mathrm{A}$ and $\mathrm{B}$. Consequently, the overall structure remains the same as that of the DC case, with the same underlying dependence on the atomic response slope $\partial \rho(\delta)/\partial \Delta S$. The main difference is the temporal integration over the finite measurement duration $T$, which replaces the instantaneous photon number $\bar{n}_0$ with the accumulated count $n_{\text{total}}$.

\section{Sensitivity Enhancement via Cavity EIT}

\begin{figure}
	\centering  
	\includegraphics[width=\linewidth]{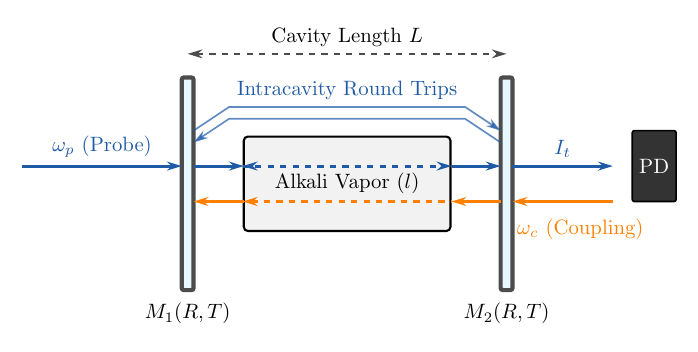}
	\caption{Schematic of the cavity EIT configuration for enhanced electric field sensing. A Rydberg atomic vapor cell of length $l$ is enclosed within a FP cavity of length $L$ composed of two partially reflecting mirrors ($M_1$ and $M_2$). The weak probe field ($\omega_p$) resonates within the cavity, experiencing multiple round-trip reflections, while the strong coupling field ($\omega_c$) establishes the EIT condition.}
	\label{fig:Cavity_sche}
\end{figure}

While the bias-assisted scheme proposed in the previous section successfully linearizes the sensor response and eliminates frequency ambiguity, the ultimate achievable sensitivity remains fundamentally constrained by the finite optical depth of the single-pass atomic cell. To overcome this limitation and further suppress the noise floor towards the fundamental photon shot-noise limit, we introduce a cavity EIT configuration where the Rydberg atomic medium is enclosed within a FP cavity, as illustrated in Fig.~\ref{fig:Cavity_sche}.

\begin{figure*}
	\centering  
	\includegraphics[width=\linewidth]{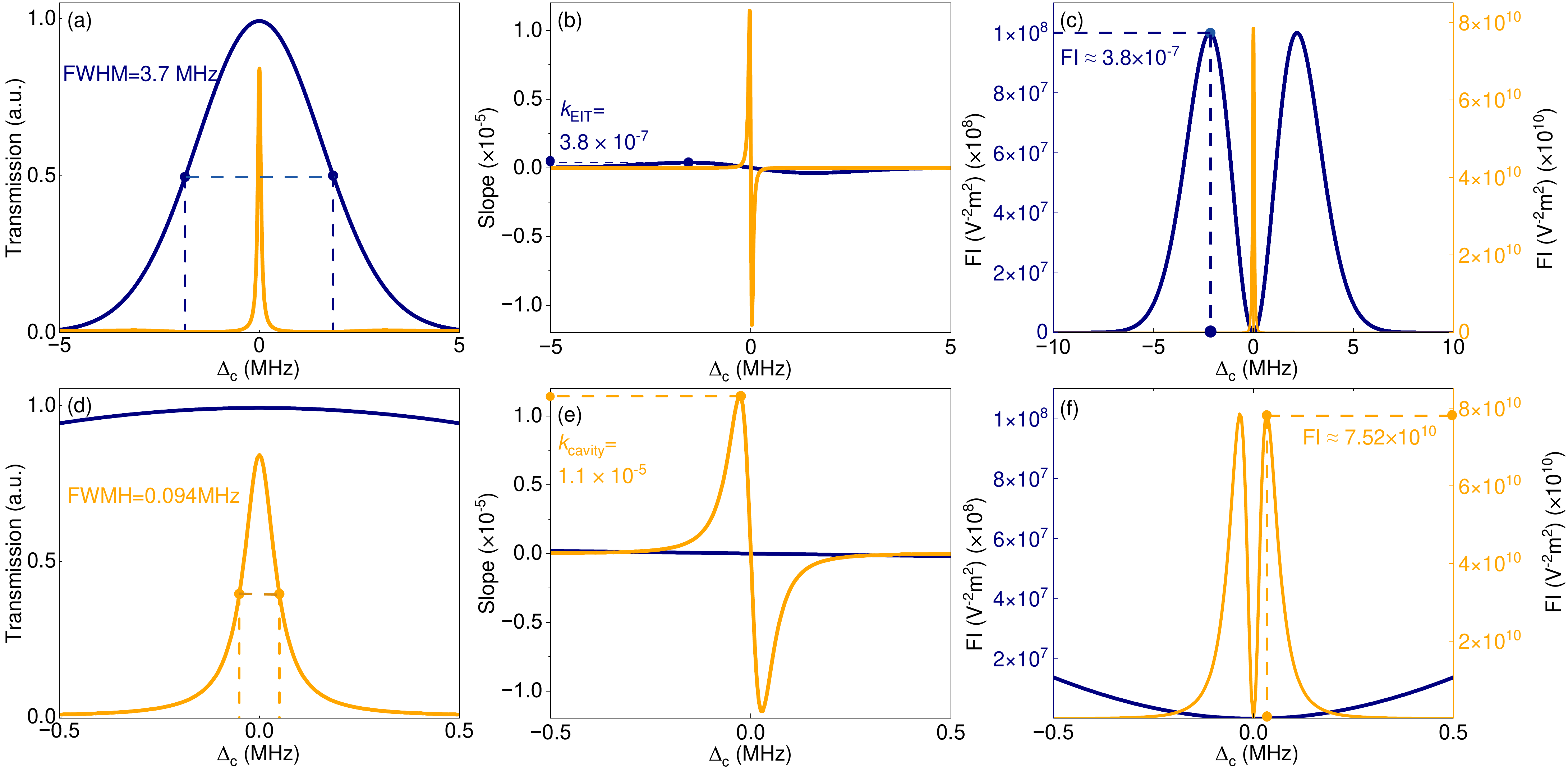}
	\caption{Performance comparison between free-space EIT and cavity-enhanced EIT. (a) Transmission spectrum of the free-space EIT scheme versus coupling detuning $\Delta_c$, with a linewidth of $\mathrm{FWHM}\approx 3.7~\mathrm{MHz}$. (b) Corresponding spectral slope, with a maximum value of $k_{\text{EIT}}\approx 3.8\times10^{-7}$. (c) Corresponding FI, with a peak value of $\approx 1\times10^{8}~\mathrm{V^{-2}m^{2}}$. (d)--(f) Enlarged views of panels (a)--(c) for the cavity-enhanced EIT case. The cavity feedback compresses the linewidth to $\mathrm{FWHM}\approx 0.094~\mathrm{MHz}$, increases the maximum slope to $k_{\text{Cavity}}\approx 1.1\times10^{-5}$, and raises the peak FI to $\approx 7.52\times10^{10}~\mathrm{V^{-2}m^{2}}$. The cavity-induced linewidth narrowing therefore leads to a much steeper spectral response and an FI enhancement of more than two orders of magnitude compared with the free-space EIT case. Cavity parameters: $R=0.9$, $L=50~\mathrm{cm}$, and $l=5~\mathrm{cm}$.}
	\label{fig:cavity}
\end{figure*}

The physical origin of the enhancement lies in the interplay between the FP cavity resonance and the atomic EIT response. For an empty cavity, the probe transmission follows the standard Airy function, determined by the mirror reflectivity $R$ (with transmissivity $T=1-R$) and the round-trip phase $\Phi$. When a Rydberg vapor cell of length $l$ is placed inside a cavity of length $L$, the cavity response is modified by the complex atomic susceptibility $\chi$~\cite{Xiao2024}. On the one hand, the imaginary part $\mathrm{Im}[\chi]$ introduces intracavity absorption, giving a round-trip attenuation factor $\kappa=\exp(-2\omega_p l\,\mathrm{Im}[\chi]/c)$ and thereby reducing the effective cavity finesse. On the other hand, the real part $\mathrm{Re}[\chi]$ contributes an additional dispersive phase shift, so that the round-trip phase is modified to $\Phi \approx (2L/c)[\Delta+(\omega_p l/2L)\,\mathrm{Re}[\chi]]$, where $\Delta=\omega_p-\omega_{\mathrm{cav}}$ denotes the probe--cavity detuning. Taking both the absorptive and dispersive effects into account, the cavity EIT transmission spectrum can be written as~\cite{Xiao2024}
\begin{equation}
	S(\omega_p) = \frac{T^2 \kappa^{1/2}}{1 + R^2\kappa^2 - 2R\kappa \cos\left[ \frac{2L}{c} \left( \Delta + \frac{\omega_p l}{2L} \mathrm{Re}[\chi] \right) \right]}.
	\label{eq:Cavity_EIT_final}
\end{equation}

To visualize the spectral narrowing and sensitivity enhancement predicted by Eq.~\eqref{eq:Cavity_EIT_final}, we perform numerical simulations comparing the free-space EIT and cavity-enhanced EIT schemes, as summarized in Fig.~\ref{fig:cavity}. Figures~\ref{fig:cavity}(a)--(c) show the transmission spectrum, spectral slope, and FI of the free-space EIT system, respectively. The free-space EIT spectrum exhibits a relatively broad transparency feature with a linewidth of $\mathrm{FWHM}\approx 3.7~\mathrm{MHz}$, a maximum slope of $k_{\mathrm{EIT}}\approx 3.8\times10^{-7}$, and a peak FI of approximately $1\times10^{8}~\mathrm{V^{-2}m^{2}}$.

\begin{figure}
	\centering  
	\includegraphics[width=\linewidth]{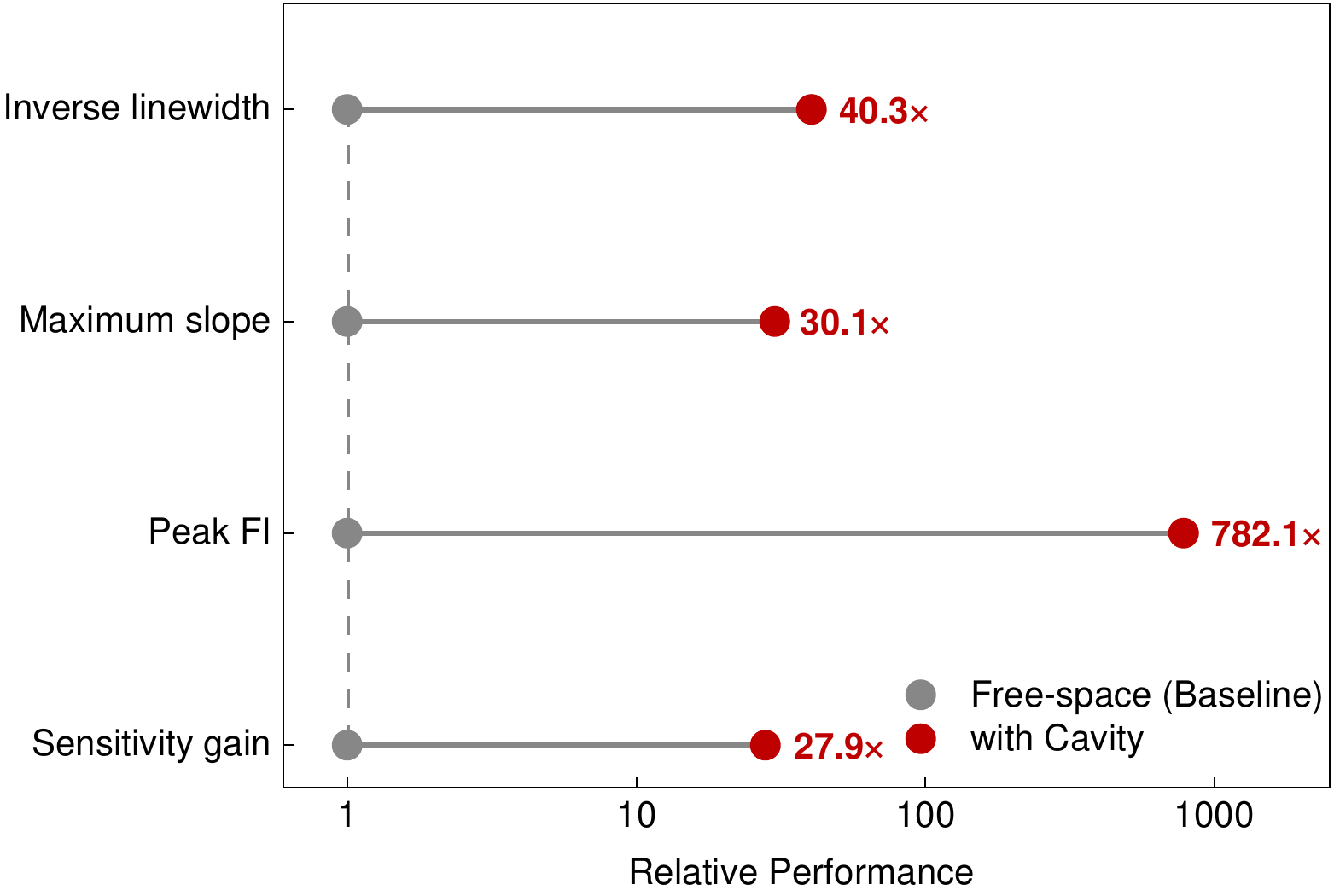}
	\caption{Summary of the performance enhancement achieved by the cavity EIT scheme relative to the free-space EIT baseline. The cavity-assisted configuration improves the inverse linewidth, maximum slope, peak FI, and sensitivity by factors of approximately $40.3$, $30.1$, $782.1$, and $27.9$, respectively, demonstrating the substantial overall gain provided by cavity enhancement.}
	\label{fig:dumbbal}
\end{figure}

Figures~\ref{fig:cavity}(d)--(f) present the corresponding enlarged results for the cavity-enhanced EIT system. Near the EIT resonance, atomic absorption is strongly suppressed, whereas the dispersive response $\mathrm{Re}[\chi]$ varies steeply with detuning. As a result, the cavity transmission is governed primarily by the susceptibility-induced phase shift rather than by intracavity loss, making the cavity resonance condition highly sensitive to small frequency variations. This effect effectively compresses the transmission feature and produces a much steeper spectral response. Consequently, the linewidth is reduced to $\mathrm{FWHM}\approx 0.094~\mathrm{MHz}$, the maximum slope increases to $k_{\mathrm{Cavity}}\approx 1.1\times10^{-5}$, and the peak FI rises to approximately $7.52\times10^{10}~\mathrm{V^{-2}m^{2}}$. As summarized in Fig.~\ref{fig:dumbbal}, relative to the free-space EIT baseline, the cavity EIT scheme improves the inverse linewidth, maximum slope, peak FI, and sensitivity by factors of approximately $40.3$, $30.1$, $782.1$, and $27.9$, respectively. These results confirm that the high-finesse cavity not only compresses the resonance linewidth, but also substantially enhances the information content and the achievable electric-field sensitivity.

However, practical deployment in power-system environments must address the high susceptibility of FP cavities to mechanical vibrations and acoustic noise. Since even sub-wavelength fluctuations in cavity length can drive the probe field out of resonance, future field-deployable designs must integrate active stabilization mechanisms, such as the mature Pound-Drever-Hall locking technique. Tightly locking the laser frequency to the cavity resonance will ensure that the theoretically predicted sensitivity enhancements are robustly maintained in harsh engineering environments.

\section{Conclusion}

In this work, we have presented a comprehensive theoretical framework and a linearized readout scheme for the quantum sensing of low-frequency electric fields using Rydberg atoms. By incorporating FI and the CRLB into the analysis of EIT spectra, we established a rigorous quantitative link between the optical transmission response and the fundamental estimation limits of the system.

To overcome the vanishing sensitivity inherent to the quadratic Stark shift in the weak-field regime, we proposed a DC-biased two-point differential measurement strategy. Our numerical simulations demonstrated that this approach effectively linearizes the sensor response for both DC and AC fields, achieving a theoretical minimum detectable field of approximately $1\times10^{-4}~\mathrm{V/m}/\sqrt{\mathrm{Hz}}$. Furthermore, we explored the integration of the atomic medium within a FP cavity. We showed that the resulting cavity EIT system significantly steepens the slope of the transmission spectrum via intracavity phase modulation, enhancing the FI by over two orders of magnitude compared to the free-space EIT scheme. These results provide theoretical guidance for the design of future Rydberg atom sensors targeting high-precision, SI-traceable monitoring of electromagnetic environments in smart grids and other low-frequency applications.

However, practical implementation in power-system environments introduces 
additional complexities. Technical noise sources---such as laser relative 
intensity noise, phase noise, and low-frequency mechanical vibrations---often 
dominate over the fundamental quantum noise and can distort the EIT spectral 
response. To bridge the gap between practical sensitivities and the 
theoretical CRLB derived in this work, advanced signal-processing strategies 
will be essential. Future implementations could incorporate advanced noise 
reduction and signal processing techniques to isolate Stark-induced spectral 
shifts from environmental noise, thereby pushing the sensor performance 
closer to its fundamental theoretical limits.

\appendix
\section{Definition of the Usable Stark-Shift Range and Origin of the Trade-Off}
\label{app:dynamic_range}

To characterize the usable linear range of the two-point differential readout, we consider the exact differential response in the Stark-shift domain,
\begin{equation}
	\rho_{\mathrm{AB}}(\Delta S;\delta)
	=
	\rho_0(\delta-\Delta S)-\rho_0(-\delta-\Delta S),
\end{equation}
where $\rho_0(\Delta_c)\equiv\rho(\Delta_c,0)$ denotes the zero-field line shape, $\Delta S$ is the Stark-induced spectral displacement, and $\delta$ is the symmetric detuning offset. In the small-shift limit, the first-order expansion of $\rho_{\mathrm{AB}}(\Delta S;\delta)$ about $\Delta S=0$ gives the local linear approximation
\begin{equation}
	\rho_{\mathrm{AB}}^{\mathrm{lin}}(\Delta S;\delta)
	=
	-2\Delta S\,\rho_0'(\delta).
\end{equation}

The departure of the exact response from the linear approximation is quantified by the relative nonlinearity
\begin{equation}
	\varepsilon_{\mathrm{nl}}(\Delta S;\delta)
	=
	\frac{
		\left|
		\rho_{\mathrm{AB}}(\Delta S;\delta)-\rho_{\mathrm{AB}}^{\mathrm{lin}}(\Delta S;\delta)
		\right|
	}{
		\left|
		\rho_{\mathrm{AB}}^{\mathrm{lin}}(\Delta S;\delta)
		\right|
	}.
\end{equation}
In this work, we adopt a tolerance $\varepsilon_0=5\%$ and define the usable Stark-shift range as
\begin{equation}
	R_{\Delta S}(\delta)
	=
	\max\left\{
	\Delta S:\,
	\varepsilon_{\mathrm{nl}}(\Delta S;\delta)\le\varepsilon_0
	\right\}.
\end{equation}
This quantity specifies the largest Stark-induced shift for which the two-point differential response can still be treated as effectively linear around the operating point $\pm\delta$.

The above definition is used to construct the range quantity discussed in the main text. For each coupling strength $\Omega_c$, the FI-optimal detuning $\delta$ is first determined from the FI analysis, and $R_{\Delta S}$ is then evaluated at $\Delta_c=\delta$, yielding the quantity $R_{\Delta S}(\Omega_c)$ plotted together with the peak FI in Fig.~\ref{fig:FI_stark}(b). Since $F_{\max}$ is governed by the transmission--slope balance in Eq.~\eqref{Eq:FI_EIT}, whereas $R_{\Delta S}$ is determined by the validity range of the local linear approximation, the two quantities generally exhibit opposite trends as $\Omega_c$ varies.

\section*{Acknowledgment} 
We thank Liang-long Wu and Dr. Dong Wei for insightful discussions and 
valuable suggestions on this work. 


\EOD

\end{document}